\documentclass[%
 reprint,
 superscriptaddress,
 amsmath,amssymb,
 aps,
 pra,
]{revtex4-2}

\usepackage{graphicx}
\usepackage{dcolumn}
\usepackage{bm}
\usepackage{mathrsfs}
\usepackage{xcolor}
\usepackage{lipsum}
\usepackage{dblfloatfix}

\newcommand{\D}{\mathrm{d}}
\def \buildrelum#1\over#2{\mathrel{\mathop{\kern0pt #2}
                     \limits_{#1}   } }

\def \widesim{~~~~\lower.40em\hbox{$\widetil$}\mskip-10mu\succ ~~~}
\def \aslim(#1) {\buildrelum #1 \over \widesim}
\def \widetil{\mathrel\mathchar"0367}

\begin{document}

\preprint{APS/123-QED}

\title{Quantum plasmonics with $N$ emitters: bright hybrid continuum selection}

\author{Georgii Semin}
\affiliation{Universit\'e Bourgogne Europe, Laboratoire Interdisciplinaire Carnot de Bourgogne ICB, UMR CNRS 6303,  21000 Dijon, France}%
\author{Hans-Rudolf Jauslin}%
\affiliation{Universit\'e Bourgogne Europe, Laboratoire Interdisciplinaire Carnot de Bourgogne ICB, UMR CNRS 6303,  21000 Dijon, France}%
\author{G\'erard Colas des Francs}
\affiliation{Universit\'e Bourgogne Europe, Laboratoire Interdisciplinaire Carnot de Bourgogne ICB, UMR CNRS 6303,  21000 Dijon, France}%
\author{St\'ephane Gu\'erin}
\affiliation{Universit\'e Bourgogne Europe, Laboratoire Interdisciplinaire Carnot de Bourgogne ICB, UMR CNRS 6303,  21000 Dijon, France}%

\date{\today}

\begin{abstract}
We construct mode-selective effective models describing the interaction of the quantum plasmon-polariton field supported by a finite dielectric medium and one or several quantum emitters.  The construction of the effective model is based on the decomposition of the field into bright modes relevant to the interaction with the emitters and dark modes, which do not interact with the emitters. We show that the quantum plasmon-polariton field can be represented equivalently by a double-continuum spectrum or by a single hybrid continuum spectrum for each emitter.
The system of the electromagnetic field coupled to a finite medium is composed of two families of continuum modes, each of them with an infinite degeneracy.  The two families are deformations of the free electromagnetic field and the free medium, induced by the interaction between them, as described by the Lippmann-Schwinger equations.
We show that if there are $N$ emitters interacting with this plasmon-polariton field, the effective interaction involves a much smaller set of bosonic continuum modes: the interacting part of the continuum can be described by $N$ non-degenerate one-dimensional continua, one for each emitter. The representation of the interaction in terms of a single hybrid continuum spectrum coincides with the one within the macroscopic Langevin model with bulk medium. This coincidence is explained by an exact compensation of two terms, one in the coupling term of the Hamiltonian and the other one in a Green tensor identity.
\end{abstract}

\maketitle



\section{\label{sec:Intro} Introduction}
 Quantum plasmonics offers new perspectives for light-matter interaction at the nanoscale \cite{Novotny2012, pitarke2007theory, Tame2013, Delga2014, Rousseaux2016, bozhevolnyi2017case, Kuisma2022, Westerberg2023, gonzalez2024light}. It has numerous applications such as single-photon sources \cite{Koenderink2009, Siampour2017, Sugawara2022}, quantum nano-optics \cite{Chang2006}, quantum plasmonic sensors \cite{Lee2021}, and polaritonic chemistry \cite{Feist2018, fregoni2022theoretical}.
 
Two approaches have been widely used for the theoretical description of quantum plasmonics for systems with an infinite homogeneous or inhomogeneous medium: a description with a medium represented by oscillators coupled to the electromagnetic field \cite{Huttner1991,Huttner1992a,Huttner1992b,Barnett1992,Hopfield1958,Wubs2001,Suttorp2004,Suttorp2004a,Wubs2004,bhat2006hamiltonian}, where the Hamiltonian is diagonalized by a technique introduced by Friedrichs \cite{friedrichs1948perturbation,friedrichs1965} and Fano \cite{fano1961effects}, and a phenomenological approach formulated in terms of a Langevin noise formalism \cite{Gruner1995,Gruner1996,Dung1998three,Scheel1998,Scheel2008,Vogel2006,Matloob1995,Matloob1997,Drezet2016,Drezet2017a,Drezet2017b}. 

The problem for a medium of finite size was formulated by canonical quantization in \cite{Dorier2019,Dorier2020}, in which the Hamiltonian diagonalization was performed using the Lippmann-Schwinger equations, showing a double continuum structure, as opposed to the framework of the macroscopic Langevin approach with an infinite bulk medium, which involves a single continuum. It was solved numerically in \cite{Na2021,Na2023} and exactly for one- \cite{Semin2024} and three-dimensional models \cite{Semin2025}. 

The Langevin model has been empirically modified by the addition of a free field component \cite{Drezet2017a,Franke2020,Na2023,Ciattoni2024,Miano2025a,Miano2025b,Miano2026}.
The modified Langevin noise model proposed in \cite{Ciattoni2024} involves two continua. It is constructed from the Heisenberg representation of the dynamics. 
From our point of view, this formulation lacks some essential aspects, like the definition of the classical Hilbert space and the ensuing bosonic Fock space with well-defined creation-annihilation operators, and the Lippmann-Schwinger equations that characterise the double continuum.
 
 In this article, we present the construction of effective (and exact) Hamiltonians characterizing the interaction of one or several quantum emitters with finite nanostructures via plasmon-polariton fields. 
It is based on the canonical quantization leading to the double continuum from the Lippmann-Schwinger equations and the complete Fock-space construction.
 We show specifically that the plasmon-polariton field can be decomposed into bright modes relevant for the interaction and dark modes, which do not interact with the emitters and are irrelevant to their dynamics. Furthermore, we show that the dark and bright mode (DBM) decomposition leads finally to an effective Hamiltonian that features a single hybrid non-degenerate continuum.
 We emphasize that the DBM decomposition is an exact result. 

 
The hybrid non-degenerate continuum Hamiltonian that we obtain
has the same form as the effective bright Hamiltonian obtained from the Langevin noise approach in \cite{hummer2013weak, Castellini2018, Feist2020}. The coincidence of these two Hamiltonians is explained by the fact that in the two-continua approach, there are two terms that cancel each other exactly, one in an intermediate formula for the couplings with the emitter and the other in the boundary term of a Green tensor identity. These two terms are missing in the Langevin Noise Approach.

The article is structured as follows. In Sec. II, we present the DBM decomposition for a single emitter. We show that the representation of the interaction in terms of a single hybrid continuum spectrum, see \eqref{Hred4} with \eqref{Omega2}, coincides with the one within the macroscopic Langevin model with bulk medium. In Sec. III
we extend the DBM decomposition to $N$ emitters. We conclude in Sec. IV.

\section{DBM decomposition for a single emitter}\label{Sect62}

\subsection{DBM decomposition: General presentation}
The DBM decomposition is an approach that constructs a mode-selective model of the interaction between quantum emitters and the electromagnetic field involving a minimal but sufficient number of modes of the EM field \cite{knoll2001coherence, Buhmann2008, Dzsotjan2010, VanVlack2012, hummer2013weak, Hakami2014, Dzsotjan2016, Rousseaux2016}. 
The DBM decomposition was first formulated for a single emitter in \cite{Buhmann2008}. It was then further developed and applied to various models in \cite{hummer2013weak, Dzsotjan2016}. The approach was generalized to the case of $N$ emitters in \cite{Castellini2018} and then applied to several applications in \cite{Varguet2019, Feist2020, Snchez-Barquilla2022, Lednev2025}, in particular using the L\"owdin orthogonalization algorithm \cite{Castellini2018, Varguet2019}, which is better suited than the Gram-Schmidt one for numerical implementation, which is known to have numerical instabilities for large $N$. In \cite{Lednev2025}, the $N$-emitter formalism was adapted to describe both emitters and detectors.
All these results were formulated within the framework of the macroscopic Langevin approach, i.e., involving a single continuum. 
In references \cite{Miano2025a, Miano2025b, Miano2026}, the DBM decomposition was partially explored in the modified Langevin noise model proposed in \cite{Ciattoni2024}. Specifically, the hybrid non-degenerate continuum Hamiltonian that we determine in this article was not obtained in \cite{Miano2025a, Miano2025b, Miano2026}.

Consider the Hamiltonian of the coupled QEs and the quantum plasmon polariton (QPP) field in the general form
\begin{equation}
 \hat H = \hat H_{\text{QE}} + \hat H_{\text{QPP}} + \hat H_{\text{int}}, 
\end{equation}
where $\hat H_{\text{QE}}$ is the free Hamiltonian of the quantum emitters, $\hat H_{\text{QPP}}$ is the free Hamiltonian of the QPP field, and $\hat H_{\text{int}}$ describes the interaction between QEs and the QPP field. One can rewrite the interaction in terms of the operators of the collective modes of the QPP field, whose set does not coincide with the full set of the field model. We call these operators {\it bright operators} and the corresponding modes, which interact with the emitters, {\it bright modes}. The remaining modes, which do not interact with the emitters and are not visible to them, are called {\it dark modes}.  The free Hamiltonian of the QPP field, decomposed on the bright and dark modes, takes the form
\begin{equation}
\label{Htot}
\hat H_{\text{QPP}}=\hat{H}^{\text{bright}}_{\text{QPP}}+\hat{H}^{\text{dark}}_{\text{QPP}}.
\end{equation}
Since we have, by construction, $[\hat H,\hat{H}^{\text{dark}}_\text{QPP}]=0$, the dark modes are not coupled with the other modes and are not affected by the interaction with the emitter, which allows one to drop the $\hat{H}^{\text{dark}}_\text{QPP}$ term in \eqref{Htot} and define the exact reduced Hamiltonian:
\begin{equation} 
\label{Hred}
\hat H_{\text{eff}}=\hat{H}^{\text{bright}}_\text{QPP}+\hat{H}_{\text{QE}}+\hat{H}^{\text{bright/QE}}_{\text{int}}
\end{equation}
with the interaction part $\hat{H}^{\text{bright/QE}}_{\text{int}}$ written in terms of the bright modes remaining equivalent to its initial form $\hat H_{\text{int}}$.

\subsection{\label{sec:Model}Plasmon-polariton quantum-emitter interaction: the canonical Hamiltonian}
\begin{figure}[t] 
   \centering
   \includegraphics[width=0.5\textwidth]{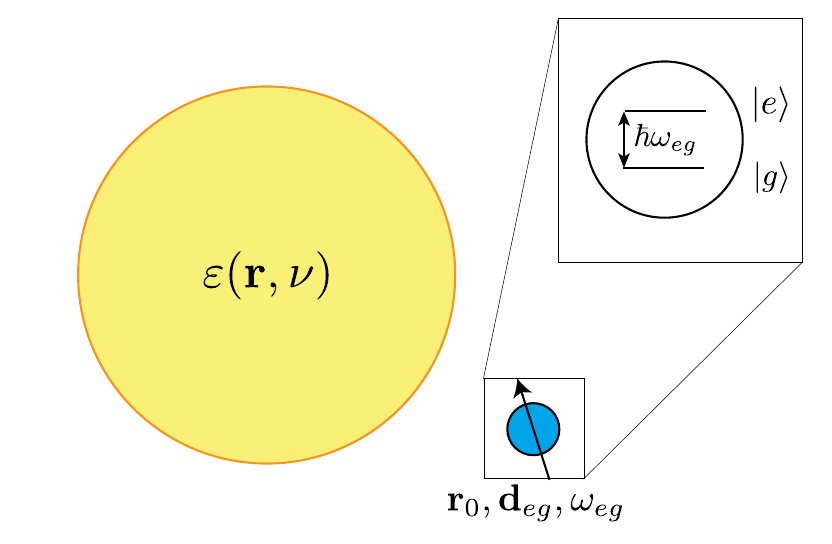} 
   \caption{System of a quantum emitter coupled with QPP supported by the dielectric structure.}
   \label{fig:scheme_1_emit}
\end{figure}

We study the model of a two-level QE, localized at a position $\mathbf{r}_0$, of states $\vert g \rangle,\vert e\rangle$ with the ground and excited energies $0$ and $\hbar\omega_{eg}$, respectively, coupled to a QPP field. We consider the three-dimensional system of the electromagnetic field coupled to a finite inhomogeneous dielectric medium of arbitrary shape. The dielectric occupies the volume $V_m$ and its properties are described by the dielectric coefficient $\varepsilon(\mathbf{x}, \nu)$ at the position $\mathbf{x} \in V_m$ inside the medium and frequency $\nu$ (see Fig.~\ref{fig:scheme_1_emit}). The QPP field results from its interaction with the electromagnetic field, and features the two types of elementary excitation operators $\hat{C}^{e\dagger}_{\kappa}$  and $\hat{C}^{m\dagger}_{\mu}\equiv \hat{C}^{m\dagger}_{\mathbf{x}, \nu, j}$, determined from the canonical diagonalization and satisfying the standard canonical commutation relations \cite{Semin2024, Semin2025}
\begin{subequations}
\begin{gather}
\bigl[ \hat{C}^e_\kappa, \hat{C}^{e\dagger}_{\kappa'} \bigr] = \delta(\kappa - \kappa'), \\
\bigl[ \hat{C}^m_\mu, \hat{C}^{m\dagger}_{\mu'} \bigr] = \delta(\mu - \mu'), \\
\bigl[ \hat{C}^e_\kappa, \hat{C}^e_{\kappa'} \bigr] =\bigl[ \hat{C}^m_\mu, \hat{C}^m_{\mu'} \bigr] =\bigl[\hat{C}^e_\kappa, \hat{C}^m_\mu\bigr]=0.
\end{gather}
\end{subequations}
They produce two types of single excitations from the vacuum, respectively:
  \begin{equation}
 \hat C^{e,\dagger} _{\kappa}\vert\varnothing\rangle = \vert\psi_{\kappa}^e\rangle,\quad \hat C^{m,\dagger} _{\mu} \vert\varnothing\rangle = \vert\psi_{\mu}^m\rangle.
\end{equation}
The corresponding Hamiltonian reads in the dipolar approximation
\begin{align}
\hat{H}=&\int \D \kappa\, \hbar\omega_{\kappa}\hat{C}^{e\dagger}_{\kappa}\hat{C}^e_{\kappa}+\int_0^{+\infty} \D \mu\, \hbar \nu_\mu\hat{C}^{m\dagger}_{\mu}\hat{C}^m_{\mu}\notag \\
&+ \hbar \omega_{eg}\hat{\sigma}_{ee} - \hat\sigma_x\otimes  \mathbf{d}_{eg} \cdot \hat{\mathbf{E}}(\mathbf{r}_0),
\label{Hcg}
\end{align}
where the integration over the multi-index $\kappa=(\mathbf{k}, \sigma_\kappa, \zeta_\kappa)$ with the wave vector $\mathbf{k}$ of norm $k = |\mathbf{k}|$, frequency $\omega_\kappa = ck$, direction of the wave vector $\mathbf{n} = \mathbf{k}/k$, polarization index $\sigma_\kappa$, and $\zeta_\kappa$ labeling the index of symmetry, i.e. associated to the real basis of generalized eigenfunctions of the differential operator $\nabla \times \nabla \times$
\begin{equation} \label{eq:basis}
	\mathbf{\Phi}_\kappa (\mathbf{r}) = (2\pi)^{-3/2}\vec{\epsilon}_{\sigma_{\kappa}} 
	\begin{cases}
	\cos(\mathbf{k}\mathbf{r}) & \zeta_\kappa = c, \\
	\sin(\mathbf{k}\mathbf{r}) & \zeta_\kappa = s.
	\end{cases}
\end{equation}
We use the abridged notation
\begin{equation}
\int \D \kappa := \int \D^3k \sum_{\sigma_\kappa = \pm} \sum_{\zeta_\kappa=c,s}.
\end{equation}
 For simplicity, we rearrange the multi-index $\kappa$ as $\kappa = (\omega_\kappa, d_\kappa)$, where we included all the degeneracy indices into $d_\kappa = (\mathbf{n}, \sigma_\kappa, \zeta_\kappa)$. The integral over $\kappa$ now denotes
\begin{align}\label{6_kappa_sum}
\int \D \kappa &= \frac{1}{c}\int_{0}^{+\infty} \!\!\!\!\D \omega_\kappa \sum_{d_\kappa} 
= \frac{1}{c}\int_{0}^{+\infty}\!\!\!\! \D \omega_\kappa \int_{|\mathbf{n}| = 1} \!\!\! \D^3 n \sum_{\sigma_\kappa} \sum_{\zeta_\kappa}.
\end{align}
The integration over the multi-index $\mu = (\mathbf{x}_\mu, \nu_\mu, j_\mu)$ with the frequency $\nu_\mu$, position inside the dielectric $\mathbf{x}_\mu \in V_m$, and vector component index $j_\mu \in \{1, 2, 3\}$ corresponds to
\begin{equation}
\label{6_mu_sum}
    \int \D\mu := \int_0^{+\infty}\! \D\nu_\mu \sum_{d_\mu}
    =
     \int_0^{+\infty} \!\D\nu_\mu \int_{V_m} \D^3x_\mu \sum_{j_\mu=1}^3,
\end{equation}
where we merged the indices $\mathbf{x}_\mu$ and $j_\mu$ into a degeneracy multiindex $d_\mu := (\mathbf{x}_\mu, j_\mu)$. The first two terms of \eqref{Hcg} represent the energy of all the QPP modes supported by the medium.
Here we have denoted the emitter operators by $\hat\sigma_{ij}= \vert i\rangle \langle j\vert$, $\hat\sigma_x= \hat{\sigma}_{eg} + \hat{\sigma}_{ge} $, and the dipole moment at the transition $e-g$ by $\mathbf{d}_{eg}$. The last term of \eqref{Hcg} is the interaction energy of the QE with the QPP electric field $\hat{\mathbf{E}}(\mathbf{r}_0)$ defined below. 

The electric field operator can be exactly decomposed in terms of creation and annihilation operators on the two sets of eigenmodes \cite{Semin2024, Semin2025}:
\begin{align}
\label{def:el_field}
\hat{\mathbf{E}}(\mathbf{r}) &= \hat{\mathbf{E}}^{e}(\mathbf{r}) + \hat{\mathbf{E}}^{m}(\mathbf{r}) \nonumber \\
&= \hat{\mathbf{E}}^{e+}(\mathbf{r}) + \hat{\mathbf{E}}^{m+}(\mathbf{r})+\text{h.c.}\nonumber\\
&=\hat{\mathbf{E}}^{+}(\mathbf{r})+\text{h.c.},
\end{align}
where the two terms of $\hat{\mathbf{E}}^{+}(\mathbf{r}) = \hat{\mathbf{E}}^{e+}(\mathbf{r}) + \hat{\mathbf{E}}^{m+}(\mathbf{r})$,
\begin{subequations}
\label{def:el_field_compme}
\begin{align}
\hat{\mathbf{E}}^{e+}(\mathbf{r}) &= -\int \D\kappa \sqrt{\frac{\hbar}{2\varepsilon_0\omega_\kappa}}\mathbf{e}_\kappa(\mathbf{r})\hat{C}^e_\kappa  \nonumber\\
&=  -\int_0^{+\infty} \frac{\D\omega_{\kappa}}{c} \sqrt{\frac{\hbar}{2\varepsilon_0\omega_{\kappa}}} \sum_{d_\kappa} \mathbf{e}_{\omega, d_\kappa}(\mathbf{r})\hat{C}^e_{\omega, d_\kappa},\\
\label{def:el_field_compm}
\hat{\mathbf{E}}^{m+}(\mathbf{r}) &= - \int \D\mu \sqrt{\frac{\hbar}{2\varepsilon_0\nu_\mu}}   \mathbf{m}_{\mu}(\mathbf{r})\hat{C}^m_\mu\notag \\
&= - \int_0^{+\infty} \D \nu \sum_{d_\mu} \sqrt{\frac{\hbar}{2\varepsilon_0\nu}}   \mathbf{m}_{\nu, d_\mu}(\mathbf{r})\hat{C}^m_{\nu, d_\mu}
\end{align}
\end{subequations}
are respectively associated by continuity with the free electromagnetic field and the medium in the uncoupling (uc) limit, $\varepsilon_i=0$, i.e. 
$\lim_{\text{uc}}\hat{\mathbf{E}}^m= 0$. The m-coefficients 
\begin{align}
\label{Gmmux-def}
\mathbf{m}_{\mathbf{x},\nu, j}(\mathbf{r}) &= -\frac{\nu^2}{c^2}  \left(\frac{2\nu\varepsilon_i(\mathbf{x},\nu)}{\pi} \right)^{1/2} \bar{\bar{G}}_{m+}(\mathbf{r}, \mathbf{x}, \nu),
\end{align}
where $\varepsilon_i(\mathbf{x},\nu)$ is the imaginary part of the dielectric coefficient,
are determined by the Sommerfeld classical Green tensor $ \bar{\bar{G}}_{m+}(\mathbf{r},\mathbf{r}',\nu)$, as the unique solution of the equation
\begin{equation}
\label{Gmmux-eq}
\left( - \frac{\nu^2}{c^2}\varepsilon(\mathbf{r}, \nu) + \nabla \times \nabla \times\right)\bar{\bar{G}}_{m+}(\mathbf{r}, \mathbf{x}, \nu) = \bar{\bar{I}}\delta(\mathbf{r} - \mathbf{x})
\end{equation}
satisfying the Sommerfeld outgoing radiation condition
 \begin{equation}
\label{SommerfeldG}
\lim_{|\mathbf{r}| \rightarrow {+\infty} }|\mathbf{r}|\left(\nabla \times - \frac{i\nu}{c}\frac{\mathbf{r}}{|\mathbf{r}|} \times \right)\bar{\bar{G}}_{m+}(\mathbf{r}, \mathbf{x}, \nu) = 0.
\end{equation}
The e-coefficients can also be expressed in terms of the Sommerfeld classical Green tensor:
\begin{align} 
\label{Lemmae}
	\mathbf{e}_\kappa(\mathbf{r}) =&  \omega_{\kappa}\mathbf{\Phi}_\kappa(\mathbf{r}) \notag \\
	&+ \int_{V_m} \D^3z\, \bar{\bar{G}}_{m+}(\mathbf{r}, \mathbf{z}, \omega_\kappa)\beta(\mathbf{z}, \omega_\kappa)\omega_{\kappa}\mathbf{\Phi}_\kappa(\mathbf{z}), 
\end{align}
where $\mathbf{\Phi}_\kappa(\mathbf{r})$ are the transverse eigenfunctions of $\nabla \times \nabla \times$, and
\begin{align}
\beta(\mathbf{z}, \omega_\kappa)= \frac{\omega_{\kappa}^2}{c^2}\left[\varepsilon_m(\mathbf{z}, \omega_\kappa) - 1\right].
\end{align}
The explicit forms of the e- and m-coefficients have been determined in the case of a metallic slab with a constant dielectric coefficient \cite{Semin2024}, for which the Green function is known \cite{Matloob1995}. For more complex structures, the Green function and tensor can be determined numerically (see, e.g., \cite{Na2023}). 

The $\mathbf{e}_\kappa(\mathbf{r})$ and $\mathbf{m}_{\mu}(\mathbf{r})$ coefficients, introduced in \cite[Eq.~(25)]{Dorier2019}, were referred to as boundary-assisted and medium-assisted modes, respectively, in \cite{Na2023}. However, these diagonal modes resulting from the eigenfunction solutions of the Lippman-Schwinger equations contain both contributions from the original (uncoupled) electromagnetic and medium Hamiltonians. A non-ambiguous more explicit interpretation is that one can connect $\mathbf{e}_\kappa(\mathbf{r})$ and $\mathbf{m}_{\mu}(\mathbf{r})$ by continuity to the free electromagnetic and medium modes in the uncoupling limit: $\lim_{\text{uc}}\mathbf{e}_{\kappa}(\mathbf{r})= \omega_{\kappa} \mathbf{\Phi}_{\kappa}(\mathbf{r})$, $\lim_{\text{uc}}\mathbf{m}_{\mu}(\mathbf{r})=0$, respectively. 

The imaginary part of the Green tensor is connected to the e-coefficients through the Green tensor LDOS identity:
\begin{align}
\label{GreenIdim1D_}
 &   \mathrm{Im}\,\bar{\bar{G}}_{m+}(\mathbf{x}, \mathbf{y}, \omega) = \frac{\pi c^2}{2 \omega^3}\int \D \kappa\, \mathbf{e}_{\kappa}(\mathbf{x}) \otimes \mathbf{e}^*_{\kappa}(\mathbf{y})\delta(\omega - \omega_\kappa) \notag \\
    &+ \frac{\omega^2}{c^2} \int_{V_m} \D^3 z\, \varepsilon_i(\mathbf{z}, \omega) \bar{\bar{G}}_{m+}(\mathbf{x},\mathbf{z}, \omega) \bar{\bar{G}}_{m+}^*(\mathbf{z}, \mathbf{y}, \omega)\notag \\
    &= \frac{\pi c}{2 \omega^3}\sum_{d_\kappa}\, \mathbf{e}_{\omega, d_\kappa}(\mathbf{x}) \otimes \mathbf{e}^*_{\omega, d_\kappa}(\mathbf{y})\notag \\
    &+ \frac{\omega^2}{c^2} \int_{V_m} \D^3 z\, \varepsilon_i(\mathbf{z}, \omega) \bar{\bar{G}}_{m+}(\mathbf{x},\mathbf{z}, \omega) \bar{\bar{G}}_{m+}^*(\mathbf{z}, \mathbf{y}, \omega).
\end{align}
It corresponds to the one used for an infinite bulk medium, but complemented with a boundary term \cite{Semin2024, Semin2025}.
It can also be rewritten via the $\mathbf{m}$ coefficients: 
\begin{align}
\label{GreenIdim1D_m}
	 \mathrm{Im}\,\bar{\bar{G}}_{m+}(\mathbf{x}, \mathbf{y}, \omega) &= \frac{\pi c^2}{2 \omega^3}\int \D \kappa\, \mathbf{e}_{\kappa}(\mathbf{x}) \otimes \mathbf{e}^*_{\kappa}(\mathbf{y})\delta(\omega - \omega_\kappa) \notag \\
&+ \frac{\pi c^2}{2 \omega^3} \int \D\mu \, \mathbf{m}_{\mu}(\mathbf{x}) \otimes \mathbf{m}^*_{\mu}(\mathbf{y})\delta(\omega - \nu_\mu) \notag \\
&=\frac{\pi c}{2 \omega^3}\sum_{d_\kappa} \mathbf{e}_{\omega, d_\kappa}(\mathbf{x}) \otimes \mathbf{e}^*_{\omega, d_\kappa}(\mathbf{y}) \notag \\
&+ \frac{\pi c^2}{2 \omega^3} \sum_{d_\mu} \mathbf{m}_{\omega, d_\mu}(\mathbf{x}) \otimes \mathbf{m}^*_{\omega, d_\mu}(\mathbf{y}),
\end{align}
where we have used the reciprocity relation $\bar{\bar{G}}(\mathbf{x}, \mathbf{y},\omega) = \bar{\bar{G}}^T(\mathbf{y}, \mathbf{x},\omega)$ since the present Green tensor satisfies the Sommerfeld boundary condition. From Eq.~\eqref{GreenIdim1D_}, one can express the term with the integral of the Green tensor as follows 
\begin{align}
\label{GreenintLDOS}
    &\frac{\omega^2}{c^2} \int_{V_m} \D^3 z\, \varepsilon_i(\mathbf{z}, \omega) \bar{\bar{G}}_{m+}(\mathbf{x},\mathbf{z}, \omega) \bar{\bar{G}}_{m+}^*(\mathbf{z}, \mathbf{y}, \omega)  \notag \\
    & =  \mathrm{Im}\,\bar{\bar{G}}_{m+}(\mathbf{x}, \mathbf{y}, \omega)  - \frac{\pi c^2}{2 \omega^3}\int \D \kappa\, \mathbf{e}_{\kappa}(\mathbf{x}) \otimes \mathbf{e}^*_{\kappa}(\mathbf{y})\delta(\omega - \omega_\kappa) \notag \\
   &  = \mathrm{Im}\,\bar{\bar{G}}_{m+}(\mathbf{x}, \mathbf{y}, \omega)  -  \frac{\pi c}{2 \omega^3}\sum_{d_\kappa}\, \mathbf{e}_{\omega, d_\kappa}(\mathbf{x}) \otimes \mathbf{e}^*_{\omega, d_\kappa}(\mathbf{y}).
\end{align}
One can rewrite the electric field part $\hat{\mathbf{E}}^{+}(\mathbf{r}_0)= \hat{\mathbf{E}}^{e+}(\mathbf{r}_0) + \hat{\mathbf{E}}^{m+}(\mathbf{r}_0)$ as
\begin{align}
\label{Eplus}
&\hat{\mathbf{E}}^{+}(\mathbf{r}_0) =  -\int_0^{+\infty} \D\omega \sqrt{\frac{\hbar}{2\varepsilon_0\omega}}\notag \\
&\times  \biggl( \frac{1}{c} \sum_{d_\kappa} \mathbf{e}_{\omega,d_\kappa}(\mathbf{r}_0)\hat{C}^e_{\omega,d_\kappa}+  \sum_{d_\mu} \mathbf{m}_{\omega, d_\mu}(\mathbf{r}_0)\hat{C}^m_{\omega, d_\mu}\biggr),
\end{align}
combining the integrals over $\D \omega_\kappa$ and $\D \nu_\mu$ into a single integral over $\D \omega$.


\subsection{DBM decomposition for the double-continuum spectrum}
With the expression of the electric-field operator \eqref{Eplus} composed of the sum of the contributions \eqref{def:el_field} of the e- and m-modes, the Hamiltonian \eqref{Hcg} can be written as a sum involving the two independent field terms:
\begin{equation}\label{HeHm}
\hat H = \hbar \omega_{eg} + \hat H^e + \hat H^m,
\end{equation}
where
\begin{subequations}
\begin{gather}
\hat H^e = \int \D \kappa\, \hbar\omega_{\kappa}\hat{C}^{e\dagger}_{\kappa}\hat{C}^e_{\kappa}  -  \hat{\sigma}_{x} \otimes \left[\mathbf{d}_{eg} \cdot\hat{\mathbf{E}}^{e+}(\mathbf{r}_0) + \text{h.c.}\right],\label{def:He} \\
\hat H^m = \int\D \mu\, \hbar \nu_\mu\hat{C}^{m\dagger}_{\mu}\hat{C}^m_{\mu} -  \hat{\sigma}_{x} \otimes  \left[\mathbf{d}_{eg} \cdot\hat{\mathbf{E}}^{m+}(\mathbf{r}_0) + \text{h.c.}\right].\label{def:Hm}
\end{gather}
\end{subequations}
We will perform the DBM decomposition in three steps. The first two steps are the DBM decompositions of the $\hat H^e$ and $\hat H^m$ operators, since they are independent, using the general procedure described in \cite{Castellini2018}. In the third step, we will combine these two DBM decompositions into a single hybrid DBM Hamiltonian, which, after discarding the dark modes, consists of a single one-dimensional non-degenerate continuum.

\subsubsection{DBM decomposition for the m-component}  
The term $-\mathbf{d}_{eg}\cdot\hat{\mathbf{E}}^{m+}(\mathbf{r}_0)$   in the interaction part $-\hat{\sigma}_{x} \otimes  \mathbf{d}_{eg}\cdot\hat{\mathbf{E}}^{m+}(\mathbf{r}_0)$ of $\hat H^m$ \eqref{def:Hm} can be written in the form:
\begin{align}
-\mathbf{d}_{eg} \cdot \hat{\mathbf{E}}^{m+}(\mathbf{r}_0)&=\hbar \int_0^{+\infty}  \D\nu\sum_{d_\mu} g^m_{\nu, d_\mu}(\mathbf{r}_0) \hat{C}_{\nu, d_\mu}^m\notag\\
&=\hbar \int_0^{+\infty}  \D\nu\,  \Omega^m_{\nu}(\mathbf{r}_0) \hat b^m_{\nu}(\mathbf{r}_0)
\label{Hint1}
\end{align}
with the \textit{bright annihilation operators} associated with the m-components defined as
\begin{subequations}
\begin{gather}
\hat b^m_{\nu}(\mathbf{r}_0)= \sum_{d_\mu} h^m_{\nu, d_\mu}(\mathbf{r}_0) \hat{C}^m_{\nu, d_\mu},\label{brightm}\\
 h^m_{\nu, d_\mu}(\mathbf{r}_0) =\frac{g^m_{\nu, d_\mu}(\mathbf{r}_0) }{\Omega^m_{\nu}(\mathbf{r}_0)} 
\end{gather}
\end{subequations}
and the coupling strengths, connected to the m-coefficients:
\begin{subequations}\label{gm_Omm}
\begin{gather}
 \label{gm0}
g^m_{\nu, d_\mu} (\mathbf{r}_0) =  \sqrt{\frac{1}{2\varepsilon_0\hbar\nu}}  \mathbf{d}_{eg} \cdot \mathbf{m}_{\nu,d_\mu}(\mathbf{r}_0),\\
\vert \Omega^m_{\nu}(\mathbf{r}_0)\vert^2 =\sum_{d_\mu} \vert g^{m}_{\nu, d_\mu}(\mathbf{r}_0) \vert^2.
\label{Omm}
\end{gather}
\end{subequations}
The above normalization has been chosen such that the bright operators $\hat b^m_{\nu}(\mathbf{r}_0)$ satisfy the standard bosonic commutation relations
\begin{equation}
\left[\hat b^m_{\nu}(\mathbf{r}_0),\hat b_{\nu'}^{m\dagger}(\mathbf{r}_0)\right]=\delta(\nu-\nu').
\end{equation}
One notices that the above coupling $\Omega^m_{\nu}(\mathbf{r}_0)$ and the bright operator $\hat b^m_{\nu}(\mathbf{r}_0)$ are both integrated over the complete medium. The phase of $\Omega^m_{\nu}(\mathbf{r}_0)$ can be chosen arbitrarily. We can choose $\Omega^m_{\nu}(\mathbf{r}_0)=\vert\Omega^m_{\nu}(\mathbf{r}_0)\vert$, i.e. real and positive.

Using the above definitions, one can rewrite the Hamiltonian \eqref{def:Hm} as
\begin{align}
\hat{H}^m=&\int \D\mu\, \hbar \nu_\mu\hat{C}^{m\dagger}_{\mu}\hat{C}^m_{\mu} \notag \\
&+ \hbar \hat{\sigma}_{x} \otimes \left[ \int_0^{+\infty}  \D\nu\,  \Omega^m_{\nu}(\mathbf{r}_0) \hat b^m_{\nu}(\mathbf{r}_0) + h.c.\right].
\label{Ha}
\end{align}
As shown below, the free QPP energy includes dark modes, i.e., those that are not involved in the coupled dynamics. We construct the corresponding dark operators $\hat{d}^m_{\nu,d_\mu}$, which will be eliminated from the dynamics,
by subtracting from the field operator $\hat{C}^m_{\nu, d_\mu}$ its projection on the subspace generated by the orthogonal set of operators $\hat{b}^m_{\nu}(\mathbf{r}_0)$, in a similar way of a step in a Gram-Schmidt orthogonalization procedure \cite{Dzsotjan2016, Rousseaux2016, Castellini2018}:
\begin{align}
\label{DarkModes}
\hat{d}^m_{\nu, d_\mu}(\mathbf{r}_0)  := & \hat{C}^m_{\nu, d_\mu} - \int^{+\infty}_0 \!\!\!\!\!\! \D \nu' [ \hat{C}^m_{\nu, d_\mu},  \hat{b}^{m\dagger}_{\nu'}(\mathbf{r}_0) ]  \hat{b}^m_{\nu'}(\mathbf{r}_0), \notag \\
=& \hat{C}^m_{\nu, d_\mu}-  \hat{b}^m_{\nu}(\mathbf{r}_0)   h^{m\ast}_{\nu, d_\mu}(\mathbf{r}_0) .
\end{align}
The dark operators satisfy the following commutation properties 
\begin{subequations}
\begin{align}
\left[\hat{b}^m_{\nu'}(\mathbf{r}_0),\hat{d}^{m\dagger}_{\nu, d_\mu}(\mathbf{r}_0)\right]&=\left[\hat{b}^m_{\nu'}(\mathbf{r}_0),\hat{d}^m_{\nu, d_\mu}(\mathbf{r}_0)\right]=0,\\
\left[\hat{d}^m_{\mathbf{x}, \nu, j}(\mathbf{r}_0),\hat{d}^{m\dagger}_{\mathbf{x}', \nu', j'}(\mathbf{r}_0)\right]  &= \delta(\nu-\nu') \delta_{jj'} \notag \\
\times\big[\delta(\mathbf{x}-\mathbf{x}') &- h^{m}_{\mathbf{x}, \nu, j}\!(\mathbf{r}_0) h^{m\ast}_{\mathbf{x}', \nu', j'}\!(\mathbf{r}_0)\big].\label{CommutDark}
\end{align}
\end{subequations}
The dark modes defined by \eqref{DarkModes} are neither orthogonal to each other nor normalized, which leads to the generalized commutation relations \eqref{CommutDark}. One could apply the orthogonalization procedure to obtain standard canonical quantization relations, but since these operators will not appear in the reduced model, there is no need to do it.

The free QPP energy associated with the medium can then be decomposed into the bright and dark terms as
\begin{align}
\sum_{d_\mu}\hat{C}^{m\dagger}_{\nu, d_\mu}&\hat{C}^m_{\nu, d_\mu}= \hat{b}_{\nu}^{m\dagger}(\mathbf{r}_0) \hat{b}^m_{\nu}(\mathbf{r}_0)+ \sum_{d_\mu} \hat{d}_{\nu, d_\mu}^{m\dagger}(\mathbf{r}_0) \hat{d}^m_{\nu, d_\mu}(\mathbf{r}_0).
\end{align}
Omitting the dark modes, we define the effective Hamiltonian from the above decomposition as formally presented in \eqref{Hred}:\\
\begin{align}
\hat{H}^m_{\text{eff}}=&\int_0^{+\infty}\!\!\!\!\!\!\D\nu \,\hbar \nu\, \hat{b}_{\nu}^{m\dagger}(\mathbf{r}_0) \hat{b}^m_{\nu}(\mathbf{r}_0)\notag \\
&+\hbar \hat{\sigma}_{x} \otimes \left[  \int_0^{+\infty} \!\!\!\D\nu\,  \Omega^m_{\nu}(\mathbf{r}_0) \hat b^m_{\nu}(\mathbf{r}_0)+\text{h.c.}\right].
\label{Hm_eff}
\end{align}

\subsubsection{DBM decomposition for the e-component}
The term $-\mathbf{d}_{eg} \cdot \hat{\mathbf{E}}^{e+}(\mathbf{r}_0)$ from the interaction part $- \hat\sigma_{x} \otimes \mathbf{d}_{eg}\cdot \hat{\mathbf{E}}^{e+}(\mathbf{r}_0)$ of $\hat H^e$ \eqref{def:He} can be written in the form
\begin{align}
-\mathbf{d}_{eg} \cdot \hat{\mathbf{E}}^{e+}(\mathbf{r}_0)&=\hbar \int_0^{+\infty}  \frac{\D\omega}{c} \sum_{d_\kappa} g^e_{\omega, d_\kappa}(\mathbf{r}_0) \hat{C}^e_{\omega, d_\kappa} \notag\\
&=\hbar\int_0^{+\infty}\D\omega\, \Omega^e_{\omega}(\mathbf{r}_0) \hat{b}^e_\omega(\mathbf{r}_0).
\label{Hint_e}
\end{align}
One can construct similarly as before a normalized {\it bright annihilation operator} for the e-component, that sums over the degeneracy indices $d_\kappa$:
\begin{subequations}
\begin{gather}
\hat b^e_{\omega}(\mathbf{r}_0)= \frac{1}{c} \sum_{d_\kappa} h^e_{\omega,d_\kappa} (\mathbf{r}_0) \hat{C}^e_{\omega, d_\kappa},\label{brighte}\\
h^e_{\omega,d_\kappa} (\mathbf{r}_0) = \frac{g^e_{\omega,d_\kappa} (\mathbf{r}_0)}{\Omega^e_{\omega}(\mathbf{r}_0) } 
\end{gather}
\end{subequations}
with the coefficient $g^{e}_{\omega,d_\kappa}(\mathbf{r}_0)$, given in \eqref{ge0}, and the coupling strength $\Omega^e_{\omega}(\mathbf{r}_0)$, chosen real and positive, such that
\begin{subequations}\label{ge_Ome}
\begin{align}
\label{ge0}
 g^e_{\kappa}(\mathbf{r}_0) &= \sqrt{\frac{1}{2\varepsilon_0\hbar\omega_\kappa}} \mathbf{d}_{eg} \cdot \mathbf{e}_\kappa(\mathbf{r}_0),\\
\vert \Omega^e_{\omega}(\mathbf{r}_0)\vert^2 &= \frac{1}{c}  \sum_{d_\kappa} \vert g^{e}_{\omega,d_\kappa} (\mathbf{r}_0)\vert^2,
\label{Ome}
\end{align}
\end{subequations}
and a dark operator
\begin{align}
\hat{d}^e_{\omega,d_\kappa}(\mathbf{r}_0) :=& \hat{C}^e_{\omega,d_\kappa} - \int^{+\infty}_0 \!\!\!\!\!\! \D \omega' [ \hat{C}^e_{\omega,d_\kappa}, \hat{b}^{e\dagger}_{\omega'}(\mathbf{r}_0)] \hat{b}^e_{\omega'}(\mathbf{r}_0) , \notag \\
=&\hat{C}^e_{\omega,d_\kappa} -  h^{e\ast}_{\omega,d_\kappa}(\mathbf{r}_0)\hat{b}^e_{\omega}(\mathbf{r}_0), 
\end{align}
such that 
\begin{subequations}
\begin{gather}
\left[\hat b^e_{\omega}(\mathbf{r}_0),\hat b_{\omega'}^{e\dagger}(\mathbf{r}_0)\right]=\delta(\omega-\omega'),\\
\left[\hat{b}^e_{\omega'}(\mathbf{r}_0),\hat{d}^{e\dagger}_{\omega,d_\kappa}(\mathbf{r}_0)\right]=\left[\hat{b}^e_{\omega'}(\mathbf{r}_0),\hat{d}^e_{\omega,d_\kappa}(\mathbf{r}_0)\right]=0.
\end{gather}
\end{subequations}
The free QPP energy associated with the free electromagnetic field can then be decomposed into the bright and dark terms as
\begin{equation}
\sum_{d_\kappa} \hat{C}^{e\dagger}_{\omega,d_\kappa}\hat{C}^e_{\omega,d_\kappa}=c \hat{b}_{\omega}^{e\dagger}(\mathbf{r}_0) \hat{b}^e_{\omega}(\mathbf{r}_0) + \sum_{d_\kappa} \hat{d}_{\omega,d_\kappa}^{e\dagger}(\mathbf{r}_0) \hat{d}^e_{\omega,d_\kappa}(\mathbf{r}_0).
\end{equation}
We obtain then the effective e-term of the Hamiltonian \eqref{def:He} omitting the uncoupled dark modes:
\begin{align}
\hat{H}^e_{\text{eff}}=&\int_0^{+\infty} \D\omega\, \hbar\omega\hat{b}^{e\dagger}_{\omega}(\mathbf{r}_0)\hat{b}^e_{\omega}(\mathbf{r}_0) \notag \\
&+\hbar \hat{\sigma}_{x} \otimes \left[ \int_0^{+\infty}\D\omega\, \Omega^e_{\omega}(\mathbf{r}_0) \hat{b}^e_\omega(\mathbf{r}_0) +\text{h.c.}\right].
\label{He_eff}
\end{align}

\subsubsection{Effective Hamiltonian with the double continuum}
After finding the components of the effective Hamiltonian related to the m-components and e-components, we can construct the effective component of the full Hamiltonian. By replacing the e- and m-terms in \eqref{HeHm} with $\hat H^e \rightarrow \hat H^e_\text{eff}$ and $\hat H^m \rightarrow \hat H^m_\text{eff}$ and substituting \eqref{Hm_eff} and \eqref{He_eff}, the effective Hamiltonian reads
\begin{align}
\hat{H}^\text{(em)}_{\text{eff}}&=\int_0^{+\infty} \D\omega\, \hbar\omega\left(\hat{b}^{e\dagger}_{\omega}\hat{b}^e_{\omega}+ \hat{b}_{\omega}^{m\dagger}\hat{b}^m_{\omega}\right)+\hbar \omega_{eg}\hat{\sigma}_{ee}\notag\\
&+\hbar \hat{\sigma}_{x} \otimes  \left[\int_0^{+\infty}\!\!\!\!\!\!\D\omega\, \left(\Omega^e_{\omega} \hat{b}^e_\omega +\Omega^m_{\omega} \hat b^m_{\omega} \right)+\text{h.c.}\right].
\label{Hred3}
\end{align}
For simplicity, we have omitted the explicit mention of the QE position $\mathbf{r}_0$ on which all the quantities depend in \eqref{brightm}, \eqref{Omm}, \eqref{brighte}, and \eqref{Ome}.

\subsection{Merging of the two bright continua into a single hybrid continuum} 
We can  now construct an effective Hamiltonian that contains a single hybrid continuum spectrum by merging the e- and m-continua, applying a secondary DBM decomposition. We define a normalized hybrid bright operator
\begin{subequations}\label{brightme}
\begin{gather}
\hat B_{\omega}=\frac{1}{\Omega_{\omega}}\left(\Omega^e_{\omega} \hat{b}_\omega^e +\Omega^m_{\omega} \hat b^m_{\omega}\right),\\
[\hat B_{\omega},\hat B_{\omega'}^{\dagger}]= \delta(\omega-\omega'),
\end{gather}
\end{subequations}
with the coupling $\Omega_{\omega}$ defined such that 
\begin{equation}\label{coupling_hyb_single}
\vert\Omega_{\omega}\vert^2=\vert \Omega^e_{\omega}\vert^2+ \vert \Omega^m_{\omega}\vert^2
\end{equation}
and a phase that can be chosen arbitrarily. The coupling is chosen real and positive:
\begin{align}
\Omega_{\omega}=\sqrt{\vert \Omega^e_{\omega}\vert^2+ \vert \Omega^m_{\omega}\vert^2}.
\label{brightC}
\end{align}
All the quantities in \eqref{brightme} and \eqref{brightC} depend on the QE position $\mathbf{r}_0$.
Due to the merging of two continua into a single hybrid continuum, we have to construct the secondary dark operators (also dependent on the position $\mathbf{r}_0$):
\begin{subequations}
\begin{align}
\hat{D}^e_{\omega} :=& \hat{b}_{\omega}^e - \int^{+\infty}_0 \!\!\!\!\!\! \D \omega' [\hat{b}_{\omega}^e, \hat{B}^\dagger_{\omega'}] \hat{B}_{\omega'}   \notag \\ 
=& \hat{b}_{\omega}^e -  \hat{B}_{\omega} \frac{\Omega^{e\ast}_{\omega}} {\Omega^{\ast}_{\omega}},\\
\hat{D}^m_{\omega} :=& \hat{b}^m_{\omega} - \int^{+\infty}_0 \!\!\!\!\!\! \D \omega'[\hat{b}_{\omega}^m, \hat{B}^\dagger_{\omega'}]\hat{B}_{\omega'} \notag \\
=& \hat{b}^m_{\omega} -  \hat{B}_{\omega} \frac{\Omega^{m\ast}_{\omega}}{\Omega^{\ast}_{\omega}},
\end{align}
\end{subequations}
which, combined with the definition \eqref{brightme}, lead to the following identity
\begin{align}
\hat{b}^{e\dagger}_{\omega}\hat{b}^e_{\omega}+ \hat{b}_{\omega}^{m\dagger} \hat{b}^m_{\omega} =  
\hat B^{\dagger}_{\omega}\hat B_{\omega}+\hat{D}^{e\dagger}_{\omega} \hat{D}^{e}_{\omega} +\hat{D}^{m\dagger}_{\omega} \hat{D}^m_{\omega}. 
\end{align}
Excluding the dark modes, the final reduced exact Hamiltonian reads then
\begin{align}
\hat{H}^{(\text{hyb})}_{\text{eff}}=&\int_0^{+\infty} \D\omega\, \hbar\omega \hat{B}^{\dagger}_{\omega}\hat{B}_{\omega}+\hbar \omega_{eg}\hat{\sigma}_{ee} \notag \\
&+\hbar \hat{\sigma}_{x} \otimes \left[ \!\int_0^{+\infty} \D\omega\, \Omega_{\omega}  \hat B_{\omega}+\text{h.c.}\right].
\label{Hred4}
\end{align}
It shows the involvement of a single effective hybrid continuum via the effective annihilation operator $\hat{B}_{\omega}$ composed by a superposition of the two original types of annihilation operators $ \hat{C}^e_{\kappa}$ and $\hat{C}^m_\mu$ via \eqref{brightme}, \eqref{brighte}, \eqref{brightm}:
\begin{equation}\label{bright_hyb_1}
\hat B_{\omega}=\frac{1}{\Omega_{\omega}}\biggl[\frac{1}{c}  \sum_{d_\kappa} g^e_{\omega,d_\kappa} (\mathbf{r}_0) \hat{C}^e_{\omega,d_\kappa}+ \sum_{d_\mu} g^m_{\nu, d_\mu}(\mathbf{r}_0) \hat{C}^m_{\omega, d_\mu} \biggr].
\end{equation}

One can rewrite the electric field part $\hat{\mathbf{E}}^{+}(\mathbf{r}_0)= \hat{\mathbf{E}}^{e+}(\mathbf{r}_0) + \hat{\mathbf{E}}^{m+}(\mathbf{r}_0)$ as
\begin{align}
&\hat{\mathbf{E}}^{+}(\mathbf{r}_0) =  -\int_0^{+\infty} \D\omega \sqrt{\frac{\hbar}{2\varepsilon_0\omega}}\notag \\
&\times  \biggl( \frac{1}{c} \sum_{d_\kappa} \mathbf{e}_{\omega,d_\kappa}(\mathbf{r}_0)\hat{C}^e_{\omega,d_\kappa}+  \sum_{d_\mu} \mathbf{m}_{\omega, d_\mu}(\mathbf{r}_0)\hat{C}^m_{\omega, d_\mu}\biggr),
\end{align}

\subsection{Direct DBM decomposition}
Instead of obtaining the effective Hamiltonian with the double continuum and merging two continua sequentially, one can avoid the intermediate steps and obtain the effective Hamiltonian with a hybrid single continuum immediately. Here, we show that we can obtain Eq. \eqref{Hred4} by  a direct approach of the DBM decomposition, starting from the Hamiltonian \eqref{Hcg}, written in terms of the creation-annihilation operators $ \hat{C}^e_{\kappa}$, $ \hat{C}^{e\dagger}_{\kappa}$, $\hat{C}_{\mu}^m$, $\hat{C}_{\mu}^{m\dagger}$, and obtainig the effective Hamiltonian of the form \eqref{Hred4}. First, we rewrite the interaction part of the Hamiltonian \eqref{Hcg}, expanding the electric-field operator by \eqref{Eplus}, as follows
\begin{align}
&- \mathbf{d}_{eg} \cdot \hat{\mathbf{E}}^{+}(\mathbf{r}_0) \notag \\
&=\hbar \int_0^{+\infty} \!\! \D\omega\biggl[ \frac{1}{c} \sum_{d_\kappa} g^e_{\omega, d_\kappa}(\mathbf{r}_0) \hat{C}^e_{\omega, d_\kappa} +\sum_{d_\mu} g^m_{\nu, d_\mu}(\mathbf{r}_0) \hat{C}_{\omega, d_\mu}^m\biggr]\notag\\
&=\hbar \int_0^{+\infty} \!\! \D\omega\,  \Omega_{\omega}\hat B_{\omega}.
\end{align}
Here, we introduced the hybrid bright operator, defined by
\begin{equation}
\hat B_{\omega}=\frac{1}{\Omega_{\omega}}\biggl[\frac{1}{c}  \sum_{d_\kappa} g^e_{\omega,d_\kappa} (\mathbf{r}_0) \hat{C}^e_{\omega,d_\kappa}+ \sum_{d_\mu} g^m_{\nu, d_\mu}(\mathbf{r}_0) \hat{C}^m_{\omega, d_\mu} \biggr],
\end{equation}
which has exactly the same form as \eqref{bright_hyb_1}. It satisfies the following commutation relation
\begin{equation}
\left[\hat B_{\omega}, \hat B^\dagger_{\omega'}\right] = \delta(\omega - \omega').
\end{equation}
The coupling $\Omega_{\omega}$ is chosen such that it satisfies the normalizing condition by the commutation relation, and the coupling reads
\begin{align}
\vert\Omega_{\omega}\vert^2=& \frac{1}{c} \sum_{d_\kappa}\vert g^{e}_{\omega, d_\kappa} (\mathbf{r}_0)\vert^2+\sum_{d_\mu} \vert g^{m}_{\omega, d_\mu} (\mathbf{r}_0) \vert^2\nonumber\\
=&\vert \Omega^e_{\omega}\vert^2+ \vert \Omega^m_{\omega}\vert^2,
\end{align}
where we used \eqref{Omm} and \eqref{Ome}. One can see that the coupling has the same form as \eqref{coupling_hyb_single}.\\
\\
Introducing the dark mode operators by
\begin{subequations}
\begin{align}
\hat{D}^e_{\omega, d_\kappa} :=& \hat{C}^e_{\omega, d_\kappa} -\int^{+\infty}_0 \!\!\!\!\!\! \D \omega' [\hat{C}^e_{\omega, d_\kappa}, \hat{B}^\dagger_{\omega'}] \hat{B}_{\omega'}   \notag \\ 
=&\hat{C}^e_{\omega, d_\kappa} -  \frac{1}{c} \frac{ g^{e*}_{\omega,d_\kappa}(\mathbf{r}_0) }{\Omega^*_\omega}\hat{B}_{\omega} ,\\
\hat{D}^m_{\omega, d_\mu} :=& \hat{C}^m_{\omega, d_\mu} - \int^{+\infty}_0 \!\!\!\!\!\! \D \omega'[\hat{C}^e_{\omega, d_\kappa}, \hat{B}^\dagger_{\omega'}]\hat{B}_{\omega'} \notag \\
=& \hat{C}^m_{\omega, d_\mu} -  \frac{ g^{m*}_{\nu, d_\mu}(\mathbf{r}_0)}{\Omega^{\ast}_{\omega}}\hat{B}_{\omega} ,
\end{align}
\end{subequations}
such that
\begin{subequations}
\begin{gather}
\left[\hat{B}_{\omega},  \hat{D}^e_{\omega', d_{\kappa'}} \right] = \left[\hat{B}_{\omega},  \hat{D}^{e\dagger}_{\omega', d_{\kappa'}} \right] = 0, \\
\left[\hat{B}_{\omega},  \hat{D}^m_{\omega', d_{\mu'}} \right] = \left[\hat{B}_{\omega},  \hat{D}^{m\dagger}_{\omega', d_{\mu'}} \right] = 0.
\end{gather}
\end{subequations}
Then, the non-interacting QPP term of the Hamiltonian can be rewritten in terms of the bright and dark mode operators as follows 
\begin{align}
&\frac{1}{c}\sum_{d_\kappa}\hat{C}^{e\dagger}_{\omega, d_\kappa} \hat{C}^e_{\omega, d_\kappa} + \sum_{d_\mu}\hat{C}^{m\dagger}_{\omega, d_\mu}\hat{C}^m_{\omega, d_\mu} \notag \\
&= \hat{B}^\dagger_{\omega} \hat{B}_{\omega}+ \frac{1}{c}\sum_{d_\kappa}\hat{D}^{e\dagger}_{\omega, d_\kappa}\hat{D}^e_{\omega, d_\kappa} + \sum_{d_\mu}\hat{D}^{m\dagger}_{\omega, d_\mu} \hat{D}^m_{\omega, d_\mu}. 
\end{align}
Omitting the dark mode operators since they do not participate in the dynamics of QE, one obtains the effective Hamiltonian \eqref{Hred4}.

\subsection{Connection to the imaginary part of the Green tensor} 

From the DBM decomposition, the effective exact coupling reads from \eqref{coupling_hyb_single}
\begin{align}\label{Omega2_preldos}
&\vert\Omega_{\omega}\vert^2 = \vert\Omega^e_{\omega} \vert^2 + \vert\Omega^m_{\omega}\vert^2,
\end{align}
where the coupling of the e-continuum can be expressed in terms of the e-coefficients, using  Eq.~\eqref{ge_Ome}, as follows
\begin{subequations}
\begin{align}\label{Ome_e}
&\vert\Omega^e_{\omega} \vert^2 = \frac{1}{c} \sum_{d_\kappa}\vert g^{e}_{\omega, d_\kappa} (\mathbf{r}_0)\vert^2 \notag \\
&= \frac{\omega^2}{\hbar\pi\varepsilon_0 c^2}   \frac{\pi c}{2\omega^3}  \sum_{d_\kappa}  \vert  \mathbf{d}_{eg}\cdot \mathbf{e}_{\omega, d_\kappa}(\mathbf{r}_0)  \vert^2 \notag \\
&=  \frac{\omega^2}{\hbar\pi\varepsilon_0 c^2}  \mathbf{d}_{eg}\cdot  \biggl[ \frac{\pi c}{2\omega^3}  \sum_{d_\kappa}   \mathbf{e}_{\omega, d_\kappa}(\mathbf{r}_0) \otimes \mathbf{e}^*_{\omega, d_\kappa}(\mathbf{r}_0)\biggr] \cdot  \mathbf{d}_{eg},
\end{align}
and the coupling of the m-continuum can be expressed in terms of the m-coefficients or in terms of the Green tensor and e-coefficients, using Eq.~\eqref{gm_Omm}, expression of the m-coefficient in terms of the Green tensor \eqref{Gmmux-def}, and expressing the integral of the Green tensor using Green tensor LDOS identity \eqref{GreenintLDOS}, as follows
\begin{align}\label{Omm_e}
&\vert\Omega^m_{\omega} \vert^2 = \sum_{d_\mu} \vert g^{m}_{\omega, d_\mu} (\mathbf{r}_0) \vert^2 \notag \\
&= \frac{\omega^2}{\hbar\pi\varepsilon_0 c^2} \biggl[\frac{\pi c^2}{2 \omega^3}  \sum_{d_\mu} |\mathbf{d}_{eg}\cdot  \mathbf{m}_{\nu, d_\mu}(\mathbf{r}_0)|^2\biggr] \notag \\
&= \frac{\omega^2}{\hbar\pi\varepsilon_0 c^2}\mathbf{d}_{eg}\cdot  \bigg[\frac{\pi c^2}{2 \omega^3} \sum_{d_\mu} \mathbf{m}_{\nu, d_\mu}(\mathbf{r}_0) \otimes \mathbf{m}^*_{\nu, d_\mu}(\mathbf{r}_0) \bigg] \cdot  \mathbf{d}_{eg} \notag \\
&= \frac{\omega^2}{\hbar\pi\varepsilon_0 c^2}\mathbf{d}_{eg}\cdot \bigg[\frac{\omega^2}{c^2} \int_{V_m} \D^3x\,  \varepsilon_i(\mathbf{x},\omega)\notag \\
&\times \bar{\bar{G}}_{m+}(\mathbf{r}_0,\mathbf{x},\omega)\bar{\bar{G}}_{m+}(\mathbf{x},\mathbf{r}_0,\omega)\bigg]\cdot  \mathbf{d}_{eg} \notag \\
&= \frac{\omega^2}{\hbar\pi\varepsilon_0 c^2}\mathbf{d}_{eg}\cdot  \bigg[  \text{Im}\,\bar{\bar{G}}_{m+}(\mathbf{r}_0,\mathbf{r}_0,\omega)\notag \\
& -  \frac{\pi c}{2\omega^3}  \sum_{d_\kappa}   \mathbf{e}_{\omega, d_\kappa}(\mathbf{r}_0) \otimes \mathbf{e}^*_{\omega, d_\kappa}(\mathbf{r}_0) \bigg]\cdot  \mathbf{d}_{eg}.
\end{align}
\end{subequations}
Then, due to the exact compensation of the term \eqref{Ome_e} with that of the same form, appearing in \eqref{Omm_e}, the effective coupling, given by the sum \eqref{Omega2_preldos}, reads
\begin{equation}
\label{Omega2}
\vert\Omega_{\omega}(\mathbf{r}_0)\vert^2= \frac{\omega^2}{\hbar\pi\varepsilon_0 c^2}\mathbf{d}_{eg} \cdot \text{Im}\,\bar{\bar{G}}_{m+}(\mathbf{r}_0,\mathbf{r}_0,\omega) \cdot \mathbf{d}_{eg}.
\end{equation}
This result, which connects the coupling to the imaginary part of the Green tensor, has been used in the literature \cite{hummer2013weak, Feist2020} where it was based on the bulk-medium approach \cite{Gruner1996, Dung1998three, Knoll2003, Scheel2008, buhmann2013dispersion}, but has not yet been justified for finite-size nanostructures. We have proved this result by combining the correct expressions containing the two e- and m-contributions of the electric field \eqref{def:el_field_compme} and the Green tensor LDOS identity \eqref{GreenIdim1D_}, including the boundary term.


\section{DBM decomposition for $N$ emitters}\label{Sect63}
\subsection{Model}
\begin{figure}[t] 
   \centering
   \includegraphics[width=0.5\textwidth]{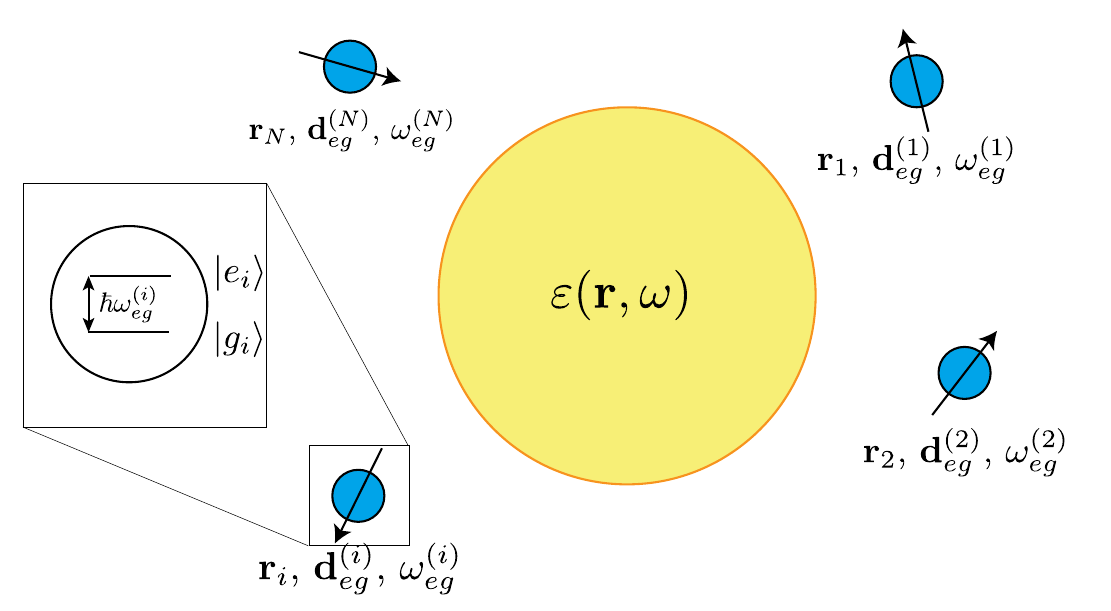} 
   \caption{System of the $N$ emitters coupled with the QPP supported by the metallic nanostructure.}
   \label{fig:scheme_N_emit}
\end{figure}

We consider a system of $N$ quantum emitters. The $k^{\text{th}}$ emitter, $k \in 1,\dots,N$, described by a two state model $|g_k\rangle$, $|e_k\rangle$ with the ground energy $0$ and excited energy $\hbar \omega^{(k)}_{eg}$ is located at the position $\mathbf{r}_k$ and has a dipole moment $\mathbf{d}^{(k)}_{eg}$. All the QEs are coupled to the QPP field supported by the metallic nanoparticle with the dielectric coefficient $\varepsilon_m(\mathbf{r}, \omega)$ (see Fig.~\ref{fig:scheme_N_emit}). For $N$ emitters, the Hamiltonian reads
\begin{align}
\hat{H}=&\int \D\kappa\, \hbar\omega_{\kappa}\hat{C}^{e\dagger}_{\kappa}\hat{C}^e_{\kappa}+\int \D\mu\, \hbar \nu_\mu\,\hat{C}^{m\dagger}_{\mu}\hat{C}^m_{\mu}  \nonumber\\
&+ \sum_{k=1}^N \hbar \omega_{eg}^{(k)}\hat{\sigma}_{ee}^{(k)} -  \sum_{k=1}^N \hat{\sigma}_{x}^{(k)} \otimes\left[\mathbf{d}_{eg}^{(k)} \cdot \hat{\mathbf{E}}^{+}(\mathbf{r}_k) +\text{h.c.}\right],
\label{H_N}
\end{align}
where the electric-field operator is given in \eqref{Eplus}. Here the operator $ \hat{\sigma}_{x}^{(k)} $ acts on the $k^{\text{th}}$ emitter.

For the case of $N$ emitters, we obtain two forms of the DBM decomposition: one with a double-continuum spectrum, where the effective Hamiltonian is expressed in terms of bright operators of the e- and m-continua, and another with a single hybrid continuum spectrum, where the effective Hamiltonian is expressed in terms of bright operators of the single continuum. In contrast to the case of a single emitter, described in Section~\ref{Sect62}, we provide the DBM decomposition for the single hybrid continuum spectrum directly from the initial Hamiltonian \eqref{H_N}, which is simpler than merging the e- and m-continua, and then compare the effective Hamiltonians for the double-continuum and single hybrid continuum spectra. Another difference from the case of a single emitter is that one must first orthonormalise the bright operators before defining the dark mode operators \cite{Castellini2018}, and we employ this orthonormalisation in both cases.
 
\subsection{DBM decomposition with double-continuum spectrum}
\subsubsection{Double-continuum bright operators}
Keeping the double-continuum structure of the Hamiltonian \eqref{H_N}, we rewrite its interaction part by substituting the electric-field operator \eqref{Eplus} 
\begin{align}
\hat{H}_\text{int} =&  -  \sum_{k=1}^N \hat{\sigma}_{x}^{(k)} \otimes \left[\mathbf{d}_{eg}^{(k)}\cdot\hat{\mathbf{E}}^{+}(\mathbf{r}_k) +\text{h.c.}\right] \nonumber\\
=& \hbar\sum_{k=1}^N \hat{\sigma}_{x}^{(k)} \otimes \bigg[\int_0^{+\infty}\!\!\!\!\!\!  \D\omega \notag \\
&\times \biggl( \sum_{d_\kappa} g^{e(k)}_{\omega, d_\kappa} \hat{C}_{\omega, d_\kappa}^e + \sum_{d_\mu} g^{m(k)}_{\omega, d_\mu} \hat{C}_{\omega, d_\mu}^m\biggr) +\text{h.c.}\biggr] \notag\\
=&\hbar\sum_{k=1}^N \hat{\sigma}_{x}^{(k)} \otimes \bigg[ \int_0^{+\infty}\!\!\!\!\!\!  \D \omega\notag \\
&\times \left( \Omega^{e(k)}_{\omega} \hat a^{e(k)}_{\omega} + \Omega^{m(k)}_{\omega} \hat a^{m(k)}_{\omega}\right) +\text{h.c.}\bigg],
\label{HaN}
\end{align}
where we introduce the \textit{bright annihilation operators} $\hat{a}^{e(k)}_{\omega}$ and $\hat{a}^{m(k)}_{\omega}$ associated with the e- and m-components respectively for each emitter, $k=1,\dots,N$, defined as
\begin{subequations}
\label{brightmek}
\begin{align}
& \hat a^{e(k)}_{\omega}= \frac{1}{c} \! \sum_{d_\kappa} h^{e(k)}_{\omega,d_\kappa}  \hat{C}^e_{\omega,d_\kappa},\label{ae_N}\\
&h^{e(k)}_{\omega,d_\kappa} =\frac{g^{e(k)}_{\omega,d_\kappa}}{\Omega^{e(k)}_{\omega}} ,\\
&\hat a^{m(k)}_{\omega}=  \sum_{d_\mu}  h^{m(k)}_{\omega, d_\mu} \hat{C}^m_{\omega, d_\mu},\label{am_N}\\
&h^{m(k)}_{\omega, d_\mu} =\frac{g^{m(k)}_{\omega, d_\mu}}{\Omega^{m(k)}_{\omega}},
\end{align}
\end{subequations}
and the individual coupling strengths,  respectively connected to the e- and m-coefficients:
\begin{subequations}
\begin{align}
\label{ge0k}
 g^{e(k)}_{\omega, d_\kappa}  &= \sqrt{\frac{1}{2\varepsilon_0\hbar\omega}} \mathbf{d}^{(k)}_{eg}\cdot \mathbf{e}_{\omega, d_\kappa}(\mathbf{r}_k) ,\\
\label{Omek}
\vert \Omega^{e(k)}_{\omega} \vert^2 &= \frac{1}{c}  \sum_{d_\kappa} \left\vert g^{e}_{\omega, d_\kappa} \right\vert^2,\\
 \label{gm0k}
g^{m(k)}_{\omega, d_\mu}  &=  \sqrt{\frac{1}{2\varepsilon_0\hbar\omega}}  \mathbf{d}^{(k)}_{eg}\cdot \mathbf{m}_{\omega, d_\mu}(\mathbf{r}_k) ,\\
\label{Ommk}
\vert \Omega^{m(k)}_{\omega} \vert^2 &=\sum_{d_\mu} \left\vert g^{m}_{\omega, d_\mu}  \right\vert^2.
\end{align}
\end{subequations}
We recall that $d_\kappa$ and $d_\mu$ are the degeneracy indices defined in \eqref{6_kappa_sum} and \eqref{6_mu_sum}.

The above normalization has been chosen such that the bright operators $\hat a^{e(k)}_{\omega}$  and $\hat a^{m(k)}_{\omega}$ satisfy the commutation relations for each emitter
\begin{subequations}
\begin{align}
&\left[\hat a^{e(i)}_{\omega},\hat a_{\omega'}^{e(j)\dagger}\right]=\delta(\omega-\omega')M^{e({ij})}_\omega,\\
&\left[\hat a^{m(i)}_{\omega},\hat a_{\omega'}^{m(j)\dagger}\right]=\delta(\omega-\omega')M^{m(ij)}_\omega
\end{align}
\end{subequations}
with
\begin{subequations}\label{overlap_em}
\begin{align}
M^{e(ij)}_\omega=&\frac{1}{c}\sum_{d_\kappa} h^{e(i)}_{\omega,d_\kappa}h^{e(j)*}_{\omega,d_\kappa}  \notag \\
=& \frac{1}{2\varepsilon_0\hbar\omega}\frac{\sum_{d_\kappa}  \mathbf{d}^{(i)}_{eg}\cdot \mathbf{e}_{\omega, d_\kappa}(\mathbf{r}_i) \otimes  \mathbf{e}^*_{\omega, d_\kappa}(\mathbf{r}_j) \cdot\mathbf{d}^{(j)}_{eg} }{\Omega^{e(i)}_{\omega}\Omega^{e(j)*}_{\omega} }, \label{modeoverlap_e}\\
M^{m(ij)}_\omega=&\sum_{d_\mu} h^{m(i)}_{\omega, d_\mu}h^{m(j)*}_{\omega, d_\mu}\notag \\
=& \frac{1}{2\varepsilon_0\hbar\omega}\frac{\sum_{d_\mu}  \mathbf{d}^{(i)}_{eg}\cdot \mathbf{m}_{\omega, d_\mu}(\mathbf{r}_i) \otimes  \mathbf{m}^*_{\omega, d_\mu}(\mathbf{r}_j) \cdot\mathbf{d}^{(j)}_{eg} }{\Omega^{m(i)}_{\omega}\Omega^{m(j)*}_{\omega} }, \label{modeoverlap_m}
\end{align}
\end{subequations}
which are standard bosonic commutation relation when $i=j$ since $  M^{e(ii)}_\omega= M^{m(ii)}_\omega=1$. 
When $i\ne j$, these commutation relations reflect the fact that the operators  $\hat a^{e(k)}_{\omega}$ and  $\hat a^{m(k)}_{\omega}$ are not orthogonal in the QE-index since, in general, $M^{e(ij)}_\omega\ne 0$ and $M^{m(ij)}_\omega\ne 0$. The $M^{e(ij)}_\omega$ and $M^{m(ij)}_\omega$ are called overlap coefficients (which are also called the elements of Gram matrices or metric matrices).

\subsubsection{Double-continuum orthogonalization}
We next construct two sets of bright operators that are mutually orthonormal by taking suitable linear combinations of the $\hat{a}^{e(i)}_{\omega}$:
\begin{subequations}
\label{nuovi}
\begin{align}
\hat{b}_{\omega}^{e(j)}= \sum_{i=1}^{N} \beta^{e(ji)}_{\omega}\hat{a}^{e(i)}_{\omega} \qquad &\text{for }j\in[1,N^{e}_{\text{ind}}],\\
\hat{b}_{\omega}^{e(j)}=0,  \qquad &\text{for } j\in[N^{e}_{\text{ind}}+1,N],
\end{align}
and for the $\hat{a}^{m(i)}_{\omega}$:
\begin{align}
\hat{b}_{\omega}^{m(j)}= \sum_{i=1}^{N} \beta^{m(ji)}_{\omega}\hat{a}^{m(i)}_{\omega}, \qquad  &\text{for }j\in[1,N^{m}_{\text{ind}}] \\
\hat{b}_{\omega}^{m(j)}=0, \qquad  &\text{for }j\in[N^{m}_{\text{ind}}+1,N]
\end{align}
\end{subequations}
where $N^{e}_{\text{ind}}\le N$, $N^{m}_{\text{ind}}\le N$  are the numbers of linearly independent operators $\hat{a}^{e(i)}_{\omega}$, $\hat{a}^{m(i)}_{\omega}$ respectively, and the coefficients $\beta^{e(ji)}_{\omega}$,  $\beta^{m(ji)}_{\omega}$ are chosen such that the new operators satisfy the orthonormality conditions, defined by the commutation relations \cite{Castellini2018} 
\begin{subequations}\label{orthcond}
\begin{align}
&\left[\hat{b}_{\omega}^{e(i)},\hat{b}_{\omega'}^{e(j)\dagger} \right] = \delta_{ij} \delta(\omega-\omega'),\\
 &\left[\hat{b}_{\omega}^{m(i)},\hat{b}_{\omega'}^{m(j)\dagger} \right] = \delta_{ij} \delta(\omega-\omega'),
\end{align}
\end{subequations}
These commutation relations imply the following relations for the $\beta^{e(ij)}_{\omega}$ and $\beta^{m(ij)}_{\omega}$ coefficients that were introduced in \eqref{nuovi}, for for $ i,j\in[1,N_{\text{ind}}]$
\begin{subequations}\label{orthcond_em_M}
\begin{align}
  &\sum_{k,l=1}^{N} \beta^{e(ik)}_{\omega}  M^{e(kl)}_\omega  \beta^{e(jl)*}_{\omega}  = \delta_{ij},\\
   &\sum_{k,l=1}^{N} \beta^{m(ik)}_{\omega}  M^{m(kl)}_\omega \beta^{m(jl)*}_{\omega}   = \delta_{ij}.
\end{align}
\end{subequations}
They can be constructed by the Gram-Schmidt method \cite{Dzsotjan2016} or by other orthonormalization procedures, such as the L\"owdin orthonormalization \cite{Castellini2018, Lwdin1950, Lwdin1956, Lwdin1998}. The explicit expression is given in Appendix~\ref{app_Lowdin}.
By inversion of \eqref{nuovi}, one can write
\begin{subequations}
\begin{align}
\hat{a}^{e(i)}_{\omega}=\sum_{j=1}^{N^{e}_{\text{ind}}}\gamma^{e(ij)}_{\omega}\hat{b}_{\omega}^{e(j)}, \quad i\in [1,N],\\
\hat{a}^{m(i)}_{\omega}=\sum_{j=1}^{N^{m}_{\text{ind}}}\gamma^{m(ij)}_{\omega}\hat{b}_{\omega}^{m(j)}, \quad i\in [1,N].
\end{align}
\label{asingolil}
\end{subequations}
The $\gamma^{e(ij)}_{\omega}$ and $\gamma^{m(ij)}_{\omega}$ coefficients are related to the $\beta^{e(ij)}_{\omega}$ and $\beta^{m(ij)}_{\omega}$ coefficients by
\begin{subequations}
\begin{align}
\gamma^{e(ij)}_{\omega} =& \sum_{l=1}^{N} M^{e(il)}_\omega \beta^{e(jl)*}_{\omega},\\
\gamma^{m(ij)}_{\omega} =& \sum_{l=1}^{N}  M^{m(il)}_\omega \beta^{m(jl)*}_{\omega},
\end{align}
\end{subequations}
and, by \eqref{orthcond_em_M}, they satisfy
\begin{subequations}
\begin{align}
 \sum_{k=1}^{N} \beta^{e(ik)}_{\omega}\gamma^{e(kj)}_{\omega}  = \delta_{i,j}\\
 \sum_{k=1}^{N}  \beta^{m(ik)}_{\omega}\gamma^{m(kj)}_{\omega}  = \delta_{i,j}.
\end{align}
\end{subequations}
The interaction term \eqref{HaN}  reads
\begin{align}
&\hat{H}^\text{(em)}_{\text{int}}=\hbar\sum_{i=1}^N \hat{\sigma}_{x}^{(i)}\notag \\
&\otimes \bigg[\int_0^{+\infty} \!\!\!\!\!\!\D\omega \sum_{j=1}^{N}\left(\chi^{e(ij)}_{\omega}\hat{b}_{\omega}^{e(j)}  +\chi^{m(ij)}_{\omega}\hat{b}_{\omega}^{m(j)} \right)+\text{h.c.}\bigg]
\label{Hint_final_}
\end{align}
with
\begin{subequations}
\begin{align}
&\chi_{\omega}^{e(ij)}=\Omega^{e(i)}_{\omega}\gamma^{e(ij)}_{\omega}=\Omega^{e(i)}_{\omega}\sum_{\ell=1}^{N} M^{e(il)}_\omega \beta^{e(jl)*}_{\omega},\\
&\chi_{\omega}^{m(ij)}=\Omega^{m(i)}_{\omega}\gamma^{m(ij)}_{\omega}=\Omega^{m(i)}_{\omega}\sum_{\ell=1}^{N} M^{m(il)}_\omega \beta^{m(jl)*}_{\omega}.
\end{align}
\label{Omijem}
\end{subequations}
We determine the dark operators by
\begin{subequations}\label{darkbme}
\begin{align}
\hat{d}_{\omega, d_\mu}^m :=& \hat{C}_{\omega, d_\mu}^{m}-\sum_{j=1}^{N^{m}_{\text{ind}}}\int^{+\infty}_0 \!\!\!\!\!\! \D \omega'[\hat{C}_{\omega, d_\mu}^{m}, \hat{b}_{\omega'}^{m(j)\dagger}] \hat{b}_{\omega'}^{m(j)}\notag \\
=& \hat{C}_{\omega, d_\mu}^{m}-\sum_{j=1}^{N^{m}_{\text{ind}}}\hat{b}_{\omega}^{m(j)}\alpha_{\omega, d_\mu}^{m(j)},\\
\hat{d}_{\omega,d_\kappa}^e :=& \hat{C}_{\omega, d_\kappa}^{e}-\sum_{j=1}^{N^{m}_{\text{ind}}}\int^{+\infty}_0 \!\!\!\!\!\! \D \omega'[ \hat{C}_{\omega, d_\kappa}^{e}, \hat{b}_{\omega'}^{e(j)\dagger}] \hat{b}_{\omega'}^{e(j)}\notag \\
=& \hat{C}_{\omega, d_\kappa}^{e}-\sum_{j=1}^{N^{e}_{\text{ind}}}\hat{b}_{\omega}^{e(j)}\alpha_{\omega,d_\kappa}^{e(j)},
\end{align}
\end{subequations}
where
\begin{subequations}
\label{coefdarkbme}
\begin{align}
\alpha_{\omega, d_\mu}^{m(j)}=\sum_{k=1}^N h^{m(k)*}_{\omega, d_\mu} \beta^{m(jk)*}_{\omega},\\
\alpha_{\omega,d_\kappa}^{e(j)}=\sum_{k=1}^N  h^{e(k)*}_{\omega,d_\kappa}\beta^{e(jk)*}_{\omega},
\end{align}
\end{subequations}
such that the non-interacting terms of QPP in \eqref{H_N},
\begin{equation}
\hat{H}_\text{QPP}=  \int_0^{+\infty}\!\!\!\!\!\! \D \omega\, \hbar\omega \biggl( \frac{1}{c}\sum_{d_\kappa}\hat{C}^{e\dagger}_{\omega,d_\kappa}\hat{C}^e_{\omega,d_\kappa} 
+ \sum_{d_\mu}\hat{C}^{m\dagger}_{\omega, d_\mu}\hat{C}^m_{\omega, d_\mu}\biggr)
\end{equation}
read
\begin{align}
\hat{H}_\text{QPP}=& \int_0^{+\infty}\!\!\!\!\!\! \D \omega \,\hbar\omega  \Biggl(  \sum_{j=1}^{N^{e}_{\text{ind}}}   \hat{b}_{\omega}^{e(j)\dagger}  \hat{b}_{\omega}^{e(j)} + \sum_{j=1}^{N^{m}_{\text{ind}}}   \hat{b}_{\omega}^{m(j)\dagger}  \hat{b}_{\omega}^{m(j)} \notag \\
& + \frac{1}{c}\sum_{d_\kappa} \hat{d}_{\omega,d_\kappa}^{e\dagger} \hat{d}_{\omega, d_\kappa}^e+ \sum_{d_\mu}\hat{d}_{\omega,d_\mu}^{m\dagger} \hat{d}_{\omega,d_\mu}^m\Biggr).
\end{align}
Excluding the dark modes, the effective Hamiltonian with the double bright continuum takes the form
\begin{align}\label{DBMHam_N_em}
\hat H^{\text{(em)}}_\text{eff} &= \int_0^{+\infty}\!\!\!\!\!\! \D \omega\, \hbar\omega  \Biggl(  \sum_{j=1}^{N^{e}_{\text{ind}}}   \hat{b}_{\omega}^{e(j)\dagger}  \hat{b}_{\omega}^{e(j)} + \sum_{j=1}^{N^{m}_{\text{ind}}}   \hat{b}_{\omega}^{m(j)\dagger}  \hat{b}_{\omega}^{m(j)}\Biggr)  \notag \\
&+ \sum_{k=1}^N \hbar \omega_{eg}^{(k)}\hat{\sigma}_{ee}^{(k)} + \hbar \sum_{i=1}^N \hat{\sigma}_{x}^{(i)} \otimes \bigg[ \int_0^{+\infty} \!\!\!\D\omega \notag \\
& \times \sum_{j=1}^{N}\left(\chi^{e(ij)}_{\omega}\hat{b}_{\omega}^{e(j)}  +\chi^{m(ij)}_{\omega}\hat{b}_{\omega}^{m(j)} \right)+\text{h.c.}\bigg].
\end{align}
We emphasize that this effective Hamiltonian is not the one that contains the least amount of the bright modes. In the following subsection, we demonstrate the derivation of a hybrid effective Hamiltonian with fewer bright modes.  

\subsection{Direct single hybrid continuum orthogonalization}
Here, we construct the effective Hamiltonian with a single hybrid continuum for each emitter, starting from the initial model with Hamiltonian \eqref{H_N}. The reason for this choice is that, for this procedure, one has to orthonormalize the bright modes only once. Alternatively, one could construct the effective Hamiltonian sequentially, first obtaining the effective Hamiltonian \eqref{DBMHam_N_em}, but in this approach, orthonormalization must be performed three times in total: twice to obtain Eq.~\eqref{DBMHam_N_em}, and once more after merging the double continua.
\subsubsection{Direct construction of a single hybrid continuum}
Instead of the orthogonalization of the bright modes operators of the e- and m-continua separately, one can introduce a bright operator of a {\it hybrid continuum}, which includes both bright operators of e- and m-continua. We rewrite the interaction part of the Hamiltonian as follows
\begin{align}\label{Hint_N_hybrid}
\hat{H}_\text{int} =& -  \sum_{k=1}^N  \hat{\sigma}_{x}^{(k)} \otimes  \left[\mathbf{d}_{eg}^{(k)} \cdot \hat{\mathbf{E}}^{+}(\mathbf{r}_k) +\text{h.c.}\right] \nonumber\\
=& \hbar\sum_{k=1}^N \hat{\sigma}_{x}^{(k)} \otimes \Bigg[ \int_0^{+\infty}\!\!\!\!\!\!  \D\omega \notag \\
& \times \Biggl( \frac{1}{c}\sum_{d_\kappa} g^{e(k)}_{\omega, d_\kappa} \hat{C}_{\omega, d_\kappa}^e + \sum_{d_\mu} g^{m(k)}_{\omega, d_\mu} \hat{C}_{\omega, d_\mu}^m\notag\Biggr) +\text{h.c.}\Biggr]\\
=& \hbar\sum_{i=1}^N  \hat{\sigma}_{x}^{(i)}\otimes \biggl(\int_0^{+\infty}\!\!\!\!\!\! \D\omega\, \Omega^{(i)}_{\omega} \hat{a}_{\omega}^{(i)}+\text{h.c.}\biggr),
\end{align}
where we defined the normalized bright operators from the interaction term of \eqref{Hint_N_hybrid}
\begin{equation}
\hat a^{(k)}_{\omega}=\frac{1}{\Omega^{(k)}_{\omega}}\Biggl[\frac{1}{c} \! \sum_{d_\kappa} g^{e(k)}_{\omega,d_\kappa}  \hat{C}^e_{\omega,d_\kappa}+\sum_{d_\mu}  g^{m(k)}_{\omega, d_\mu} \hat{C}^m_{\omega, d_\mu}\Biggr]
\label{brightmek_}
\end{equation}
satisfying the commutation relation
\begin{align}
 \left[\hat a^{(i)}_{\omega},\hat a^{(j)\dagger}_{\omega'}\right]= M^{(ij)}_\omega  \delta(\omega-\omega') 
\end{align}
with the overlap coefficient
\begin{align}\label{modeoverlap_hyb}
& M^{(ij)}_\omega= \frac{\frac{1}{c}\sum_{d_\kappa}g^{e(i)}_{\omega, d_\kappa}g^{e(j)*}_{\omega,d_\kappa}  + \sum_{d_\mu}g^{m(i)}_{\omega, d_\mu}g^{m(j)*}_{\omega, d_\mu}}{\Omega^{(i)}_{\omega}\Omega^{(j)*}_{\omega}}
\end{align}
characterizing the non-orthogonality of the operators $\hat a^{(i)}_{\omega}$ associated to the $i$-th emitter, and the coupling $\Omega^{(k)}_{\omega}$ defined through the normalization condition 
\begin{equation}
\bigl\vert\Omega^{(k)}_{\omega}\bigr\vert^2=\frac{1}{c}\sum_{d_\kappa}\Bigl\vert g^{e(k)}_{\omega, d_\kappa}\Bigr\vert^2+ \sum_{d_\mu}\Bigl\vert g^{m(k)}_{\omega, d_\mu}\Bigr\vert^2,
\end{equation}
with a phase that can be chosen arbitrarily. The coupling is chosen to be real and positive:
\begin{align}
\Omega^{(k)}_{\omega}&=\sqrt{\frac{1}{c}\sum_{d_\kappa}\left\vert g^{e(k)}_{\omega, d_\kappa}\right\vert^2+ \sum_{d_\mu}\left\vert g^{m(k)}_{\omega, d_\mu}\right\vert^2}.
\label{brightCk}
\end{align}
To calculate the overlap coefficients and the couplings more explicitly, we notice that the following sum can be expressed in terms of the imaginary part of the Green tensor using the LDOS Green tensor identity \eqref{GreenIdim1D_m}:
\begin{align}
\frac{1}{c}&\sum_{d_\kappa}g^{e(i)}_{\omega, d_\kappa}g^{e(j)*}_{\omega,d_\kappa}  + \sum_{d_\mu}g^{m(i)}_{\omega, d_\mu}g^{m(j)*}_{\omega, d_\mu}\notag \\
& =\frac{1}{2\varepsilon_0\hbar\omega}  \Biggl[\frac{1}{c}\sum_{d_\kappa}\mathbf{d}^{(i)}_{eg} \cdot\mathbf{e}_{\kappa}(\mathbf{r}_i ) \otimes \mathbf{e}^*_{\kappa}(\mathbf{r}_j)\cdot \mathbf{d}^{(j)}_{eg} \notag \\
&\qquad\qquad+\sum_{d_\mu}\mathbf{d}^{(i)}_{eg} \cdot\mathbf{m}_{\mu}(\mathbf{r}_i ) \otimes \mathbf{m}^*_{\mu}(\mathbf{r}_j) \cdot \mathbf{d}^{(j)}_{eg} \Biggr]\notag \\
&=\frac{\omega^2}{ \hbar \pi \varepsilon_0 c^2}\mathbf{d}^{(i)}_{eg} \cdot \text{Im}\,\bar{\bar{G}}_{m+}(\mathbf{r}_i ,\mathbf{r}_j,\omega)\cdot \mathbf{d}^{(j)}_{eg}.
\end{align}
Then, substituting the latter expression into \eqref{modeoverlap_hyb} and \eqref{brightCk}, we get the overlap coefficients and the couplings in terms of the imaginary part of the Green tensor:
\begin{align}\label{modeoverlap_hyb2}
M^{(ij)}_{\omega} =  \frac{\omega^2}{ \hbar \pi \varepsilon_0 c^2}\frac{\mathbf{d}^{(i)}_{eg} \cdot \text{Im}\,\bar{\bar{G}}_{m+}(\mathbf{r}_i ,\mathbf{r}_j,\omega)\cdot \mathbf{d}^{(j)}_{eg}  }{ \Omega^{(i)}_{\omega}\Omega^{(j)*}_{\omega}}
\end{align}
\begin{align}
\Omega^{(k)}_{\omega} = \sqrt{\frac{\omega^2}{\hbar\pi\varepsilon_0c^2}\mathbf{d}^{(k)}_{eg} \cdot \text{Im}\,\bar{\bar{G}}_{m+}(\mathbf{r}_k,\mathbf{r}_k,\omega) \cdot \mathbf{d}^{(k)}_{eg}}.
\label{brightCk}
\end{align}
{\bf Remark:} One can express the hybrid bright operator $\hat a^{(k)}_\omega$ in terms of the unorthogonalized bright operators of the e- and m-continua $\hat{a}^{e(k)}_\omega$ and $\hat{a}^{m(k)}_\omega$ of Eqs.~\eqref{ae_N} and \eqref{am_N} as follows
\begin{equation}\label{ahyb_ae_am}
\hat a^{(k)}_{\omega}=\frac{1}{\Omega^{(k)}_{\omega}}\left[\Omega^{e(k)}_{\omega}\hat{a}^{e(k)}_\omega +\Omega^{m(k)}_{\omega} \hat a^{m(k)}_{\omega}\right].
\end{equation}
The overlap coefficient can also be expressed in terms of the overlap coefficients of the e- and m-modes, given in \eqref{overlap_em}, as follows
\begin{align}
& M^{(ij)}_\omega= \frac{\Omega^{e(i)}_{\omega}\Omega^{e(j)*}_{\omega}M^{e(ij)}_\omega  + \Omega^{m(i)}_{\omega}\Omega^{m(j)*}_{\omega'}M^{m(ij)}_\omega }{\Omega^{(i)}_{\omega}\Omega^{(j)*}_{\omega}}.
\end{align}
Finally, one can express the coupling $\Omega^{(k)}_\omega$ in terms of the couplings $\Omega^{e(k)}_\omega$ and $\Omega^{m(k)}_\omega$, given by \eqref{Omek} and \eqref{Ommk}, as follows
\begin{equation}\label{Om_hyb_em}
\bigl|\Omega^{(k)}_\omega \bigr|^2 = \bigl|\Omega^{e(k)}_\omega \bigr|^2 + \bigl |\Omega^{m(k)}_\omega\bigr |^2.
\end{equation}

\subsubsection{Orthogonalization of the hybrid bright operators}
We next construct a set of bright operators that are mutually orthonormal by taking suitable linear combinations of the $\hat{a}_{\omega}$:
\begin{align}
\hat{B}_{\omega}^{( j )}= \sum_{i=1}^{N} \beta^{(ji)}_{\omega}\hat{a}^{(i)}_{\omega}, \qquad& j\in[1,N_{\text{ind}}],\notag \\
\hat{B}_{\omega}^{(  j )}=0, \qquad& j\in[N_{\text{ind}}+1,N],
\label{nuovi_}
\end{align}
where $N_{\text{ind}}\le N$ is the number of linearly independent operators $\hat{a}^{(i)}_{\omega}(\mathbf{r}_i)$ and the coefficients $\beta^{(ji)}_{\omega}$ are chosen such that the new operators satisfy the orthonormality condition 
\begin{align}\label{orthcond}
\left[\hat{B}_{\omega}^{(i)},\hat{B}_{\omega'}^{(j)\dagger} \right]& = \delta_{ij} \delta(\omega-\omega'),
\end{align}
which expressed in terms of the $\beta^{(ij)}_{\omega} $ read, for $ i,j\in[1,N_{\text{ind}}]$,
\begin{align}
  \sum_{k,l=1}^{N} \beta^{(ik)}_{\omega}  M^{(kl)}_\omega \beta^{(jl)*}_{\omega} = \delta_{ij}.
\end{align}
They can be constructed by the Gram-Schmidt method \cite{Dzsotjan2016} or by other orthonormalization procedures, such as the L\"owdin orthonormalization \cite{Castellini2018, Lwdin1950, Lwdin1956, Lwdin1998}, as described in Appendix~\ref{app_Lowdin}.
By inversion of \eqref{nuovi_}, one can write
\begin{equation}
\hat{a}^{(i)}_{\omega}=\sum_{j=1}^{N_{\text{ind}}}\gamma^{(ij)}_{\omega}\hat{B}_{\omega}^{(j)}, \quad i\in [1,N],
\label{asingolil_}
\end{equation}
where $\gamma^{(ij)}_{\omega}$ is defined as
\begin{equation}\label{gamma_hyb}
 \gamma^{(ij)}_{\omega} =  \sum_{l=1}^{N}  \beta^{(jl)*}_{\omega}  M^{(il)}_\omega
\end{equation}
that, by \eqref{asingolil_}, satisfies
\begin{equation}\label{betagamma_delta}
 \sum_{k=1}^{N}  \beta^{(ik)}_{\omega}\gamma^{(kj)}_{\omega}  = \delta_{ij}.
 \end{equation}
The interaction term of \eqref{Hint_N_hybrid} can be then written as
\begin{equation}
\hat{H}^\text{(hyb)}_{\text{int}}=\hbar\sum_{i=1}^N \hat{\sigma}_{x}^{(i)}\otimes\left[\int_0^{+\infty} \!\!\!\!\!\!\D \omega\sum_{j=1}^{N_{\text{ind}}}  \chi_{\omega}^{(ij)} \hat{B}_{\omega}^{(j)}+\text{h.c.}\right]
\label{Hint_final}
\end{equation}
with
\begin{equation}
\chi_{\omega}^{(ij)}=\Omega^{(i)}_{\omega}\gamma^{(ij)}_{\omega}=\Omega^{(i)}_{\omega}\sum_{l=1}^{N}  M_{\omega}^{(il)} \beta^{(jl)*}_{\omega}.
\label{Omij}
\end{equation}
We determine the corresponding dark operators as
\begin{subequations}
\begin{align}
\hat{D}_{\omega,d_\kappa}^e&:= \hat{C}_{\omega, d_\kappa}^{e}-\sum_{j=1}^{N_{\text{ind}}}\int^{+\infty}_0 \!\!\!\!\!\! \D \omega'[\hat{C}_{\omega, d_\kappa}^{e}, \hat{B}_{\omega'}^{(j)\dagger}]\hat{B}_{\omega'}^{(j)}\notag \\
&= \hat{C}_{\omega, d_\kappa}^{e}-\sum_{j=1}^{N_{\text{ind}}}\hat{b}_{\omega}^{(j)}\alpha_{\omega,d_\kappa}^{e(j)},\\
\hat{D}_{\omega, d_\mu}^m&:= \hat{C}_{\omega,d_\mu}^{m}-\sum_{j=1}^{N_{\text{ind}}} \int^{+\infty}_0 \!\!\!\!\!\! \D \omega'[\hat{C}_{\omega,d_\mu}^{m}, \hat{B}_{\omega'}^{(j)\dagger}] \hat{B}_{\omega'}^{(j)}\notag \\
&= \hat{C}_{\omega,d_\mu}^{m}-\sum_{j=1}^{N_{\text{ind}}}\hat{B}_{\omega}^{(j)}\alpha_{\omega, d_\mu}^{m(j)},
\end{align}
\label{darkbme_}
\end{subequations}
with
\begin{subequations}
\label{coefdarkbme_}
\begin{align}
\alpha_{\omega,d_\kappa}^{e(i)}=&\sum_k h^{e*}_{\omega, d_\kappa}\beta^{(ik)*}_{\omega}\frac{\Omega^{e(k)*}_{\omega}}{\Omega^{(k)*}_{\omega}},\\
\alpha_{\omega, d_\mu}^{m(i)}=&\sum_k h^{m*}_{\omega, d_\mu} \beta^{(ik)*}_{\omega} \frac{ \Omega^{m*}_{\omega}}{ \Omega^{*}_{\omega}},
\end{align}
\end{subequations}
 such that the non-interacting terms
\begin{equation}
\hat{H}_\text{QPP}=  \int_0^{+\infty}\!\!\!\!\!\! \D \omega\, \hbar\omega  \biggl( \frac{ 1}{c}\sum_{d_\kappa}\hat{C}^{e\dagger}_{\omega,d_\kappa}\hat{C}^e_{\omega,d_\kappa}+\sum_{d_\mu}\hat{C}^{m\dagger}_{\omega, d_\mu}\hat{C}^m_{\omega,d_\mu}\biggr)
\end{equation}
are written as
\begin{align}
&\hat{H}_\text{QPP}=\int_0^{+\infty}\!\!\!\!\!\! \D \omega\, \hbar\omega  \Bigg( \sum_{j=1}^{N_{\text{ind}}}   \hat{B}_{\omega}^{(j)\dagger}  \hat{B}_{\omega}^{(j)} \notag \\
&+ \frac{ 1}{c}\sum_{d_\kappa}\hat{D}^{e\dagger}_{\omega,d_\kappa}\hat{D}^e_{\omega,d_\kappa}+ \sum_{d_\mu}\hat{D}_{\omega, d_\mu}^{m\dagger} \hat{D}_{\omega, d_\mu}^m\Bigg).
\end{align}
Excluding the dark modes, the effective Hamiltonian takes the form\\
\begin{align}\label{DBMHam_N_hyb}
&\hat H^\text{(hyb)}_{\text{eff}} =  \sum_{j=1}^{N_{\text{ind}}}  \int_0^{+\infty} \D \omega\, \hbar\omega \hat{B}_{\omega}^{(j)\dagger}  \hat{B}_{\omega}^{(j)} + \sum_{k=1}^N \hbar \omega_{eg}^{(k)}\hat{\sigma}_{ee}^{(k)} \notag \\
& + \hbar\sum_{i=1}^N \hat{\sigma}_{x}^{(i)}\otimes \biggl(  \int_0^{+\infty} \D\omega\sum_{j=1}^{N_{\text{ind}}}  \chi_{\omega}^{(ij)} \hat{B}_{\omega}^{(j)}+\text{h.c.}\biggr). 
\end{align}
This effective Hamiltonian is the same one as the one we would obtain by merging the two orthogonalized continua of \eqref{DBMHam_N_em} and eliminating the dark modes appearing in the merging, as was done explicitly for a single emitter (see Section II.C).

\section{\label{sec:Conclusions}Conclusions}
In this article, we have derived exact mode-selective effective models of the QPP field coupled to one or several quantum emitters using the dark and bright mode decomposition of the three-dimensional quantum plasmonic-polaritonic field. The construction of the bright modes operators for a system with several emitters requires the orthonormalisation of the bright operators, which one can provide using, for example, Gram-Schmidt or L\"{o}wdin methods.

We constructed the effective models in two ways: by preserving the two-continua structure of the QPP spectrum and by introducing a single hybrid continuum. The coupling between the quantum emitters and the hybrid bright modes of the QPP is given by the Green tensor of the medium, which follows from the Green tensor LDOS identity, proven in Refs.~\cite{Semin2024, Semin2025}. Finally, we showed that one can always represent the two-continua DBM decomposition as the hybrid continuum decomposition. The resulting hybrid non-degenerate continuum Hamiltonians, see  \eqref{Hred4} and \eqref{DBMHam_N_hyb} for a single emitter and $N$ emitters, respectively, 
have the same form as the effective bright Hamiltonians obtained from the Langevin noise approach in \cite{hummer2013weak, Castellini2018, Feist2020}. This result can be seen as a justification of its use.


This construction shows that the subset of effectively interacting models can be described by $N$ one-dimensional, nondegenerate continuum bosonic modes, one for each emitter. We emphasize that the effective Hamiltonians with a single hybrid continuum contain fewer bright modes compared to the double-continuum effective Hamiltonians, which makes the former a preferable choice for numerical analysis due to lower computational resources required. Once the Hamiltonian is written in this form, one can apply some methods to select the dominant modes to construct simple approximations involving only finite-dimensional matrices, like e.g., quasi-normal modes \cite{sauvan2013theory, franke2019quantization, Ge2015, Hughes2018}, or other \cite{Dzsotjan2016, Rousseaux2016, Castellini2018, Varguet2019, Feist2020}.
\section*{Acknowledgements}
We acknowledge support from the EUR-EIPHI Graduate School (17-EURE-0002), the QuanTEdu-France project (ANR-22-CMAS-0001), the Conseil R\'egional de Bourgogne-Franche-Comt\'e, and the Dijon M\'etropole.

\begin{appendix}

\section{L\"{o}wdin orthogonalization for the bright operators}\label{app_Lowdin}

This Appendix is based on the L\"{o}wdin orthogonalization presented in \cite{Castellini2018}.

The generalization of the construction of the effective Hamiltonians is not direct for $N > 1$ emitters since the set of operators $\hat a^{(i)}_\omega$, defined in Eqs.~\eqref{ae_N}, \eqref{am_N}, or \eqref{brightmek_}, is not orthogonal, and one can see it through the commutation relation
\begin{equation}\label{orthog_general}
[\hat{a}^{(i)}_{\omega}, \hat{a}^{(i)\dagger}_{\omega}] = \delta(\omega - \omega')M^{(ij)}_\omega
\end{equation}
with the overlap coefficient $M^{(ij)}_\omega$. To solve this problem, we construct a set of bright operators that are mutually orthonormal by taking suitable linear combinations of the $\hat a^{(i)}_\omega$:
\begin{equation}\label{app_bright}
\hat{b}_{\omega}^{(j)}=\sum_{i=1}^N \beta_{\omega}^{(ji)} \hat{a}^{(i)}_{\omega}, \quad j=1, \ldots, N_{\mathrm{ind}}
\end{equation}
where $N_{ind} \leq N$ is the number of linearly independent operators $\hat a^{(i)}_\omega$, and the coefficients $\beta_\omega^{(ij)}$ are chosen such that the new operators satisfy the orthonormality condition:
\begin{equation}
\begin{aligned}
{\left[\hat{b}_{\omega}^{(i)}, \hat{b}_{\omega'}^{(j) \dagger}\right] } & = \delta(\omega-\omega^{\prime}) \delta_{i j}, \\
{\left[\hat{b}_{\omega}^{(i)}, \hat{b}_{\omega^{\prime}}^{(j)}\right] } & =0 .
\end{aligned}
\end{equation}
For this purpose, we discuss the L\"{o}wdin orthogonalization method below.

In order to implement the orthonormalization procedure, we have to identify the number of linearly independent operators $\hat a^{(i)}_\omega$. In \cite{Dzsotjan2016}, it has been shown that if $|M_\omega^{(ij)}| = 1$ with $i > j$, then it means that $\hat{a}^{(i)}_{\omega}$ and $\hat{a}^{(j)}_{\omega}$ overlap completely, i.e. the are the same operators up to a phase factor. Multiplying the commutation relation \eqref{orthog_general} by $M_\omega^{(ij)*}$ leads to
\begin{equation}
\hat{a}^{(i)}_{\omega}=M_{\omega}^{(ij)} \hat{a}^{(j)}_{\omega}.
\end{equation}
Thus, we can identify a pair of linearly dependent field operators. There is also the possibility that one operator can be expressed as a linear combination of two or more other operators. To identify the number $N_\text{ind}$, we define for the $N$ bright operators $\hat a^{(i)}_\omega$, the $N \times N$ overlap matrix $M_\omega$ via the overlap coefficients $M_\omega^{(ij)}$:
\begin{equation}
M_{\omega}:=\begin{pmatrix}
1 & M_\omega^{(2,1)} & \cdots & M_\omega^{(N, 1)} \\
M_\omega^{(1, 2)} & \ddots & \ddots & \vdots \\
\vdots & \ddots & \ddots & M_\omega^{(N, N-1)} \\
M_\omega^{(1, N)} & \cdots & M_\omega^{(N-1, N)} & 1
\end{pmatrix}.
\end{equation}
The overlap matrix $M_\omega$ is Hermitian and positive semi-definite, so we can diagonalize it through a $N \times N$ unitary matrix $T_\omega$:
\begin{align}
 D_{\omega} & := \text{diag}\left(\lambda_{1, \omega}, \ldots, \lambda_{N_{\mathrm{ind}}, \omega}, 0, \ldots, 0\right) \\
& = T_{\omega}^{\dagger} M_{\omega} T_{\omega}.
\end{align}
We implement Löwdin’s canonical orthonormalization, obtaining a new set of bright operators
\begin{equation}\label{Lowdin}
\mathbf{B}_{\omega}=\mathbf{A}_{\omega} T_{\omega} D^{-1/2}_{\omega}
\end{equation}
with
\begin{equation}
D^{-1/2}_{\omega}=\operatorname{diag}\left(\lambda_{1, \omega}^{-1/2}, \ldots, \lambda_{N_{\text {ind}}, \omega}^{-1/2}, 0, \ldots, 0\right)
\end{equation}
through the one-row arrays
\begin{align}
\mathbf{B}_{\omega} & :=\left(\hat{b}_{\omega}^{(1)}, \ldots, \hat{b}_{\omega}^{\left(N_{\text{ind}}\right)}, 0, \ldots, 0\right), \\
\mathbf{A}_{\omega} & :=\left(\hat{a}^{(1)}_{\omega}, \ldots, \hat{a}^{(N)}_{\omega}\right).
\end{align}
Substituting the expressions of the matrices in \eqref{Lowdin}, each bright operator $\hat{b}_{\omega}^{(i)}$ of the orthogonal set can be expressed in terms of the old bright operators as follows
\begin{subequations}
\begin{align}
\hat{b}_{\omega}^{(j)}= \lambda_{j,\omega}^{-1 / 2} \sum_{i=1}^N T_{\omega}^{(ij)} \hat{a}^{(i)}_{\omega},& \qquad  j \in\left[1, N_{\mathrm{ind}}\right], \\ 
\hat{b}_{\omega}^{(j)} = 0,& \qquad j \in\left[N_{\mathrm{ind}}+1, N\right],
\end{align}
\end{subequations}
where $T_{\omega}^{(ij)} $ are the elements of $T_{\omega}$. Thus, one can define the orthogonalization coefficients
\begin{equation}\label{beta_Lowdin}
\beta_{\omega}^{(ji)}:=\lambda_{j,\omega}^{-1 / 2} T_{\omega}^{(ij)},
\end{equation}
which give the linear combination in \eqref{app_bright}.

Compared with the Gram-Schmidt orthogonalization, the L\"{o}wdin orthogonalization provides simple, compact formulas and an algorithm that is more stable in numerical implementations \cite{TrefethenBau}, allowing, in principle, the treatment of a large number of emitters. This can be seen by inverting \eqref{Lowdin},
\begin{equation}
\mathbf{A}_{\omega}=\mathbf{B}_{\omega} D^{1/2}_{\omega} T_{\omega},
\end{equation}
which has the form of a singular value decomposition, which has good numerical stability properties.

\end{appendix}

\end{document}